\documentclass[english]{report}
\usepackage[T1]{fontenc}
\usepackage[latin9]{inputenc}
\usepackage{babel}

\usepackage{verbatim}
\usepackage{amsmath}
\usepackage{graphicx}
\usepackage{setspace}
\usepackage{amssymb}
\usepackage{esint}
\usepackage[unicode=true, pdfusetitle,
 bookmarks=true,bookmarksnumbered=false,bookmarksopen=false,
 breaklinks=false,pdfborder={0 0 1},backref=false,colorlinks=false]
 {hyperref}

\makeatletter

\newcommand{\lyxmathsym}[1]{\ifmmode\begingroup\def\b@ld{bold}
  \text{\ifx\math@version\b@ld\bfseries\fi#1}\endgroup\else#1\fi}

\makeatother

\begin{document}
\thispagestyle{empty}

\begin{center}
{\large \includegraphics[bb=250bp 0bp 200bp 0bp]{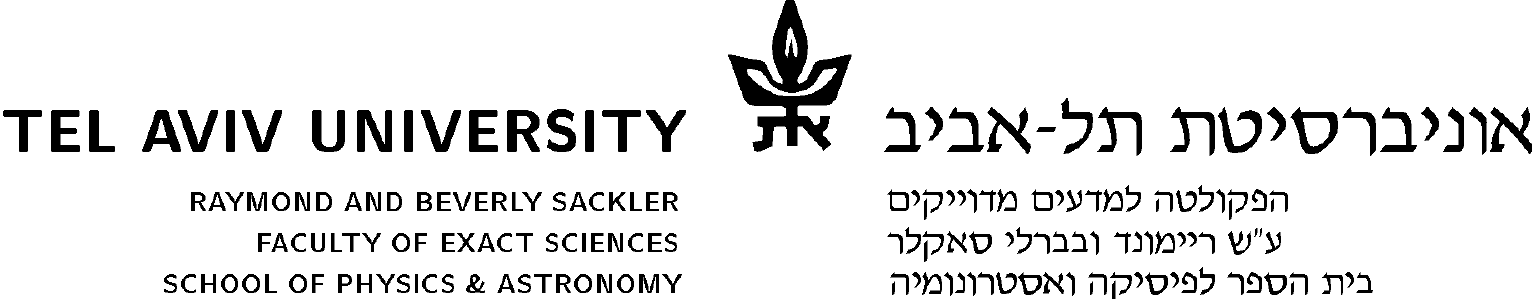}}
\par\end{center}{\large \par}

{\large \vspace{0.5cm}
}{\large \par}

\begin{center}
{\huge Relativistic Holographic Hydrodynamics}
\par\end{center}{\huge \par}

{\large \vspace{0.125cm}
}{\large \par}

\begin{center}
{\huge from Black Hole Horizons}
\par\end{center}{\huge \par}

{\large \vspace{0.5cm}
}{\large \par}

\begin{singlespace}
\begin{center}
{\large A thesis submitted in partial fulfillment}
\par\end{center}{\large \par}

\begin{center}
{\large of the requirements for the degree of}
\par\end{center}{\large \par}

\begin{center}
{\large Master in Science}
\par\end{center}{\large \par}
\end{singlespace}

{\large \vspace{0.5cm}
}{\large \par}

\begin{singlespace}
\begin{center}
{\large At Tel-Aviv University}
\par\end{center}{\large \par}

\begin{center}
{\large School of Physics and Astronomy}
\par\end{center}{\large \par}
\end{singlespace}

{\large \vspace{0.75cm}
}{\large \par}

\begin{singlespace}
\begin{center}
By
\par\end{center}

\begin{center}
{\large Adiel Meyer}
\par\end{center}{\large \par}
\end{singlespace}

{\large \vspace{0.75cm}
}{\large \par}

\begin{singlespace}
\begin{center}
Under the supervision of
\par\end{center}

\begin{center}
{\large Prof. Yaron Oz}
\par\end{center}{\large \par}
\end{singlespace}

{\large \vspace{1cm}
}{\large \par}

\begin{center}
March 2011
\par\end{center}

{\large \pagebreak{}}{\large \par}

\pagenumbering{roman} 

\begin{center}
\textbf{\LARGE Abstract}
\par\end{center}{\LARGE \par}

{\large \bigskip{}
\bigskip{}
}{\large \par}

{\large We consider the AdS/CFT correspondence in the hydrodynamic
regime up to the second order in a derivative expansion. We demonstrate
that the fluid conservation equations are equivalent to Einstein's
constraint equations projected on different hyper-surfaces. We derive
that result for hyper-surfaces of the form $r=R\left(x^{\alpha}\right)$
up to the first order in a derivative expansion of the metric. At
the second order expansion, we introduce the notion of different black
hole horizons, and focus on two particular horizon hyper-surfaces:
the event horizon and the apparent horizon. We calculate the temperature
and entropy current for the apparent horizon and show that the latter
agrees with the area increase theorem for the black hole, and differs
from the entropy current calculated for the event horizon.}{\large \par}

{\large \tableofcontents{}}{\large \par}

\chapter{{\LARGE Introduction}}

\pagenumbering{arabic} 

{\large The AdS/CFT correspondence proposed in \cite{key-21}, plays
a significant role in understanding strongly coupled conformal gauge
field theories. One important application of this duality is the study
of the effective description of these strongly coupled gauge field
theories in the long wavelength regime}%
\footnote{We are actually using the small Knudsen number regime, which is explained
in the following chapters.%
}{\large . This effective description is the hydrodynamic plasma description.
The correspondence states that a strongly coupled conformal field
theory corresponds to a weakly coupled string theory, and the latter,
in a certain regime, reduces to the Einstein classical field equations
with a negative cosmological constant in an asymptotically Anti de
Sitter space-time. This correspondence is suitable for describing
out-of equilibrium fluid flow (the hydrodynamic description) which
is dual to out-of-equilibrium long wavelength dynamics of a black
brane. An example for this is the fluid-gravity correspondence presented
in \cite{key-3}. The AdS/CFT duality has a remarkable application
in describing strongly coupled QCD type plasma \cite{key-23}.}{\large \par}

{\large The equations governing relativistic hydrodynamics are conservation
equations, namely the relativistic navier-stokes equations \cite{key-7,key-6}.
Those equations can be found from parts of Einstein's equations in
the dual gravitational description \cite{key-2,key-22}. In this thesis
we will project those equations on different hyper-surfaces and verify
that every hyper-surface reproduce the same conservation equations.}{\large \par}

{\large From thermodynamics we get the notion of many thermodynamic
quantities that can be generalized for hydrodynamics, for example:
temperature, chemical potential and entropy current. The latter will
be dual to the area of a hyper-surface that admits the area increase
theorem. In this thesis we will explore the different hyper-surfaces
and we will focus on two of them, namely: the event and apparent horizons.}{\large \par}

{\large We will start by describing the relativistic hydrodynamic
fluid and the derivative expansion in chapter 2. At the end of the
chapter we will explain the Weyl formalism, which we will work with
in the rest of the thesis. In chapter 3 we will set the stage for
the gravity description and in chapter 4 we will provide basic geometric
quantities. At the end of chapter 4 we will present parts of Einstein's
equation which are dual to the conservation equations of the hydrodynamic
fluid, and in the following two chapter 5, 6 we will explore those
equations for different hyper-surfaces. For this we will introduce
the apparent horizon hyper-surface in chapter 6. In chapter 7 we will
calculate the entropy current and the temperature for the apparent
horizon, and we will present our conclusion in chapter 8.}{\large \par}

{\large \newpage{}}{\large \par}

\chapter{{\LARGE Relativistic Hydrodynamics }}

{\large Hydrodynamics \cite{key-6} is a theoretical model which describes
the behavior of fluids in motion. It treats the fluid as having local
domains which are in equilibrium, hence one can define in those domains
definite thermodynamic quantities such as temperature, pressure, entropy,
etc. }{\large \par}

{\large One can address a fluid as a relativistic only if the macroscopic
velocity of the fluid is close to the speed of light.}{\large \par}

\section{{\large Hydrodynamics equations}}

{\large In order to derive the relativistic equations of fluid dynamics
we first need to derive the form of the energy momentum }\emph{\large stress
tensor}{\large .}{\large \par}

{\large If we look on a $d$ dimensional fluid element}%
\footnote{Can be understood as a volume within the fluid which contains a large
number of molecules%
}{\large{} in its }\emph{\large local rest frame,}{\large{} we will realize
that in that frame Pascal law is valid, which means that the pressures
on the different surfaces surrounding it are equal. Therefore, we
get $T^{ii}=p$. On that local frame the component $T^{00}$ of the
stress tensor is just the local internal energy density of the fluid
$\varepsilon$. The other components of the stress tensor are zero
$T^{0i}=0$.}{\large \par}

{\large If we introduce the $d$-velocity of the fluid $u^{\mu}$,
then in the local rest frame of the fluid we get $u^{0}=-1$ and $u^{i}=0$
. Then the energy momentum stress tensor of an ideal fluid will be:}{\large \par}

{\large \begin{equation}
T_{Ideal}^{\mu\nu}=\varepsilon u^{\mu}u^{\nu}+p\left(\eta^{\mu\nu}+u^{\mu}u^{\nu}\right)\label{eq:Ideal}\end{equation}
}{\large \par}

{\large If the fluid equation of state is provided, then one can find
the connection between the pressure and the energy density.}{\large \par}

{\large From the definition of the $d$-velocity in the local rest
frame, we can calculate its norm (which will be valid in any frame):}{\large \par}

{\large \begin{equation}
u^{\mu}u_{\mu}=-1\label{eq:normalization}\end{equation}
}{\large \par}

{\large We can define the tensor which multiplies the pressure on
(\ref{eq:Ideal}) as the }\emph{\large projector tensor}{\large{} $P^{\mu\nu}$,
because it projects along the transverse direction of the fluid velocity.}{\large \par}

{\large We would like to consider a fluid which undergoes viscosity
effects, so we will need to introduce a viscous stress tensor $\tau^{\mu\nu}$
, which will be added to the ideal stress tensor:}{\large \par}

{\large \begin{equation}
T^{\mu\nu}=T_{Ideal}^{\mu\nu}+\tau^{\mu\nu}\end{equation}
}{\large \par}

{\large We have some freedom in the determination of the viscous stress
tensor that we would like to set. First, we will consider the viscous
stress tensor to be proportional to the derivatives of $u^{\mu}\left(x\right)$.
Second, we will need to consider how viscosity will affect the fluid
in its local rest frame. In order to do so we will require that the
energy density and the momentum densities in the local rest frame
will not change due to viscous effects, which means $\tau^{00}=0$,
$\tau^{0i}=0$, and because the fluid velocity $u^{i}=0$, we have
in the local rest frame of the fluid the following expression:}{\large \par}

{\large \begin{equation}
\tau^{\mu\nu}u_{\nu}=0\end{equation}
}{\large \par}

{\large This result is correct not only in the rest frame of the fluid
but also in any Lorentz frame, this result is called the }\emph{\large Landau\textemdash{}Lifshitz
frame}{\large . }{\large \par}

{\large We will also specify the equations of motion that can be written
in a simple form, because they are just the conservation equations
of the stress tensor:}{\large \par}

{\large \begin{equation}
\partial_{\mu}T^{\mu\nu}=0\label{eq:the equations of motion}\end{equation}
}{\large \par}

\section{{\large Hydrodynamics as an effective Conformal Field Theory}}

{\large When working in the regime of small Knudsen number (long wavelength)
the relativistic field theory has a relativistic hydrodynamic description
\cite{key-7}. }{\large \par}

{\large The Knudsen number is just the correlation length of the fluid
$\ell_{cor}$ divided by the characteristic length scale $L$ of the
variations of the macroscopic fields, so we can write the condition
of effective hydrodynamics description as: }{\large \par}

{\large \begin{equation}
Kn=\frac{\ell_{cor}}{L}\ll1\label{eq:expansion requirement}\end{equation}
}{\large \par}

{\large Under this condition we can expand the stress tensor in a
small parameter $Kn\ll1$ and get:}{\large \par}

{\large \begin{equation}
T^{\mu\nu}\left(x\right)=\sum_{l=0}^{\infty}T_{l}^{\mu\nu}\left(x\right),\quad\quad T_{l}^{\mu\nu}\sim\left(Kn\right)^{l}\label{eq:stress tensor expansion}\end{equation}
where $T_{l}^{\mu\nu}\left(x\right)$ is determined locally by the
value of the velocity $u^{\mu}\left(x\right)$, the pressure $p\left(x\right)$}%
\footnote{We are assuming that the equation of state has been given.%
}{\large{} and their derivatives. For example the zeroth order will
be the ideal with no derivatives.}{\large \par}

{\large If we consider a relativistic Conformal Field Theory with
a finite temperature $T$, we get: \begin{equation}
T_{\mu}^{\mu}=0\label{eq:conformality requirement}\end{equation}
}{\large \par}

{\large Therefore, we can find the equation of state from this condition
and we get $p=\frac{\varepsilon}{d-1}$ . From dimensional analysis
we have $p=aT^{d}$ and $\varepsilon=\left(d-1\right)aT^{d}$ where
$a$ is a normalization coefficient. Then equation (\ref{eq:Ideal})
takes the form:}{\large \par}

{\large \begin{equation}
T_{Ideal}^{\mu\nu}=aT^{d}\left(\eta^{\mu\nu}+du^{\mu}u^{\nu}\right)\end{equation}
}{\large \par}

{\large In order to find the viscous stress tensor $\tau^{\mu\nu}$,
we will use the conformality requirement (\ref{eq:conformality requirement}),
from which we will see that the viscous stress tensor has to be traceless,
i.e., $\tau_{\mu}^{\mu}=0$. In addition, if we apply the equations
of motion (\ref{eq:the equations of motion}) to the zeroth order
stress tensor, we get a connection between the derivatives of the
velocity of the fluid $u^{\mu}\left(x\right)$ to the derivatives
of the fluid temperature $T\left(x\right)$, so we can replace, in
the next order, the derivatives of $T\left(x\right)$ with the derivatives
of $u^{\mu}\left(x\right)$. We can follow this procedure to all higher
order in this iterative form, in order to eliminate completely the
derivatives of $T\left(x\right)$.}{\large \par}

{\large By those two conditions and the symmetry of the stress tensor,
one can find the stress tensor by taking all possible terms with different
coefficients\cite{key-8}. Here we will present the complete stress
tensor to second order in derivatives expansion,}{\large \par}

{\large \begin{equation}
T^{\mu\nu}=aT^{d}\left(\eta^{\mu\nu}+du^{\mu}u^{\nu}\right)-2\eta\sigma^{\mu\nu}+\eta\tau_{\Pi}\Sigma_{\left(0\right)}^{\mu\nu}+\lambda_{1}\Sigma_{\left(1\right)}^{\mu\nu}+\lambda_{2}\Sigma_{\left(2\right)}^{\mu\nu}+\lambda_{3}\Sigma_{\left(3\right)}^{\mu\nu}\label{eq:stress tensor for conformal fluid 2nd order}\end{equation}
where the shear tensor is defined by:}{\large \par}

{\large \begin{equation}
\sigma_{\mu\nu}=P_{\mu}^{\alpha}P_{\nu}^{\beta}\partial_{(\alpha}u_{\beta)}-\frac{\partial_{\alpha}u^{\alpha}}{d-1}P_{\mu\nu}\end{equation}
the vorticity tensor is defined by:}{\large \par}

{\large \begin{equation}
\omega_{\mu\nu}=P_{\mu}^{\alpha}P_{\nu}^{\beta}\partial_{[\alpha}u_{\beta]}\end{equation}
and}{\large \par}

{\large \[
\Sigma_{\left(0\right)}^{\mu\nu}=2P^{\mu\alpha}P^{\nu\beta}u^{\lambda}\partial_{\lambda}\sigma_{\alpha\beta}+2\frac{\partial_{\alpha}u^{\alpha}}{d-1}\sigma^{\mu\nu},\quad\Sigma_{\left(1\right)}^{\mu\nu}=4\sigma_{\lambda}^{\mu}\sigma^{\nu\lambda}-4\frac{\sigma_{\alpha\beta}\sigma^{\alpha\beta}}{d-1}P^{\mu\nu},\]
\begin{equation}
\Sigma_{\left(2\right)}^{\mu\nu}=2\sigma^{\mu\lambda}\omega_{\lambda}^{\nu}+2\sigma^{\nu\lambda}\omega_{\lambda}^{\mu},\quad\Sigma_{\left(3\right)}^{\mu\nu}=\omega_{\lambda}^{\mu}\omega^{\nu\lambda}-\frac{\omega_{\alpha\beta}\omega^{\alpha\beta}}{d-1}P^{\mu\nu}.\end{equation}
}{\large \par}

{\large The brackets represents symmetric or anti-symmetric tensors
$A_{(\alpha\beta)}=\frac{1}{2}\left(A_{\alpha\beta}+A_{\beta\alpha}\right),\quad$
$A_{[\alpha\beta]}=\frac{1}{2}\left(A_{\alpha\beta}-A_{\beta\alpha}\right)$.}{\large \par}

{\large The coefficients $\eta,\tau_{\Pi},\lambda_{1},\lambda_{2},\lambda_{3}$
are called }\emph{\large transport coefficients}{\large , and in the
literature \cite{key-6} $\eta$ is also called the }\emph{\large shear
viscosity}{\large{} coefficient. }{\large \par}

\section{{\large Weyl covariant formulation}}

{\large In this section, we will introduce the Weyl covariant formulation
as done in \cite{key-14}, which is useful for conformal fluid.}{\large \par}

{\large We will consider a $d>3$ conformal fluid in a curved background
with metric $g_{\mu\nu}$}%
\footnote{In the rest of the thesis we will consider a flat background metric
$\eta_{\mu\nu}$, thus for a flat background one needs to replace
$g_{\mu\nu}\rightarrow\eta_{\mu\nu}$ and $\nabla_{\lambda}\rightarrow\partial_{\lambda}$.%
}{\large{} and we will look at some observables of the fluid. We will
consider a conformal transformation:}{\large \par}

{\large \begin{equation}
g_{\mu\nu}=e^{2\phi}\widetilde{g}_{\mu\nu},\quad g^{\mu\nu}=e^{-2\phi}\widetilde{g}^{\mu\nu}\label{eq:conformal transformation of the metric}\end{equation}
where $\phi=\phi\left(x^{\mu}\right)$.}{\large \par}

{\large As in the previous section we will denote by $u^{\mu}$ the
$d$-velocity of the fluid with the normalization (\ref{eq:normalization}).
From this normalization and (\ref{eq:conformal transformation of the metric})
we can find the transformation rule of the $d$-velocity by: $g_{\mu\nu}u^{\mu}u^{\nu}=\widetilde{g}_{\mu\nu}\widetilde{u}^{\mu}\widetilde{u}^{\nu}=-1$
and we get: $u^{\mu}=e^{-\phi}\widetilde{u}^{\mu}$.}{\large \par}

{\large However, if we look on the transformation rule of the covariant
derivative of $u^{\mu}$ we get:}{\large \par}

{\large \begin{equation}
\nabla_{\mu}u^{\nu}=e^{-\phi}\left(\widetilde{\nabla}_{\mu}\widetilde{u}^{\nu}+\delta_{\mu}^{\nu}\widetilde{u}^{\sigma}\partial_{\sigma}\phi-\widetilde{g}_{\mu\lambda}\widetilde{u}^{\lambda}\widetilde{g}^{\nu\sigma}\partial_{\sigma}\phi\right)\end{equation}
which does not transform homogeneously. In order to deal with homogeneous
transformations, which are allowed in a conformal fluid, we define
a }\emph{\large Weyl covariant derivative}{\large{} $D_{\lambda}$
in the following manner: if a tensorial quantity $Q_{\nu\cdots}^{\mu\cdots}$
transforms $Q_{\nu\cdots}^{\mu\cdots}=e^{-\omega\phi}\widetilde{Q}_{\nu\cdots}^{\mu\cdots}$
then the weyl covariant derivative will transform:$D_{\lambda}Q_{\nu\cdots}^{\mu\cdots}=e^{-\omega\phi}\widetilde{D}_{\lambda}\widetilde{Q}_{\nu\cdots}^{\mu\cdots}$,
where:}{\large \par}

{\large \begin{eqnarray}
D_{\lambda}Q_{\nu\cdots}^{\mu\cdots} & \equiv & \nabla_{\lambda}Q_{\nu\cdots}^{\mu\cdots}+\omega A_{\lambda}Q_{\nu\cdots}^{\mu\cdots}\nonumber \\
 &  & +\left[g_{\lambda\alpha}A^{\mu}-\delta_{\lambda}^{\mu}A_{\alpha}-\delta_{\alpha}^{\mu}A_{\lambda}\right]Q_{\nu\cdots}^{\alpha\cdots}+\ldots\label{eq:Weyl covariant derivative}\\
 &  & -\left[g_{\lambda\nu}a^{\alpha}-\delta_{\lambda}^{\alpha}A_{\nu}-\delta_{\nu}^{\alpha}\right]Q_{\alpha\cdots}^{\mu\cdots}-\ldots\nonumber \end{eqnarray}
}{\large \par}

{\large Where $A_{\lambda}$ is a one-form that can be determined
uniquely by the requirements that the Weyl covariant derivative of
the $d$-velocity will be traceless and transverse to the fluid direction,
i.e., $D_{\mu}u^{\mu}=0$ and $u^{\lambda}D_{\lambda}u^{\mu}=0$.
We get:}{\large \par}

{\large \begin{equation}
A_{\nu}=u^{\lambda}\nabla_{\lambda}u_{\nu}-\frac{\nabla_{\lambda}u^{\lambda}}{d-1}u_{\nu}\end{equation}
}{\large \par}

{\large Note that the Weyl covariant derivative is metric compatible,
i.e: $D_{\lambda}g_{\mu\nu}=0$.}{\large \par}

{\large We will present here additional transformations of some observables
of the fluid:}{\large \par}

{\large \begin{eqnarray}
D_{\mu}u^{\nu} & = & \nabla_{\mu}u^{\nu}+u_{\mu}u^{\lambda}\nabla_{\lambda}u^{\nu}-\frac{\nabla_{\lambda}u^{\lambda}}{d-1}P_{\mu}^{\nu}\nonumber \\
 & = & \sigma_{\mu}^{\nu}+\omega_{\mu}^{\nu}=e^{-\phi}\widetilde{D}_{\mu}\widetilde{u}^{\nu},\\
\sigma^{\mu\nu} & \equiv & \frac{1}{2}\left(P^{\mu\lambda}\nabla_{\lambda}u^{\nu}+P^{\nu\lambda}\nabla_{\lambda}u^{\mu}\right)-\frac{\nabla_{\lambda}u^{\lambda}}{d-1}P_{\mu}^{\nu}\nonumber \\
 & = & \frac{1}{2}\left(D^{\mu}u^{\nu}+D^{\nu}u^{\mu}\right)=e^{-3\phi}\widetilde{\sigma}^{\mu\nu},\\
\omega^{\mu\nu} & \equiv & \frac{1}{2}\left(P^{\mu\lambda}\nabla_{\lambda}u^{\nu}-P^{\nu\lambda}\nabla_{\lambda}u^{\mu}\right)\nonumber \\
 & = & \frac{1}{2}\left(D^{\mu}u^{\nu}-D^{\nu}u^{\mu}\right)=e^{-3\phi}\widetilde{\omega}^{\mu\nu}\end{eqnarray}
}{\large \par}

{\large \begin{equation}
D_{\lambda}\sigma_{\:\mu}^{\lambda}=\left(\nabla_{\lambda}-\left(d-1\right)A_{\lambda}\right)\sigma_{\:\mu}^{\lambda}\end{equation}
}{\large \par}

{\large \begin{equation}
D_{\lambda}\omega_{\mu}^{\:\lambda}=\left(\nabla_{\lambda}-\left(d-3\right)A_{\lambda}\right)\omega_{\mu}^{\:\lambda}\end{equation}
}{\large \par}

{\large \begin{equation}
u^{\lambda}D_{\lambda}\sigma_{\mu\nu}=P_{\mu}^{\:\alpha}P_{\nu}^{\:\beta}\nabla_{\lambda}\sigma_{\alpha\beta}+\frac{\nabla_{\lambda}u^{\lambda}}{d-1}\sigma_{\mu\nu}\end{equation}
}{\large \par}

{\large We will also define a }\emph{\large Weyl covariant Riemann
curvature tensor}{\large{} for a vector field with a weight $\omega$
,i.e.$V^{\mu}=e^{-\omega\phi}\widetilde{V}^{\mu}$ by}{\large \par}

{\large \begin{equation}
\left[D_{\mu},D_{\nu}\right]V_{\lambda}=\omega\mathcal{F_{\mu\nu}}V_{\lambda}+\mathcal{R_{\mu\nu\lambda}^{\quad\,\alpha}}V_{\alpha}\end{equation}
}{\large \par}

{\large with}{\large \par}

{\large \begin{equation}
\mathcal{F_{\mu\nu}}=\nabla_{\mu}A_{\nu}-\nabla_{\nu}A_{\mu}\end{equation}
}{\large \par}

{\large \begin{equation}
\mathcal{R_{\mu\nu\lambda\sigma}}=R_{\mu\nu\lambda\sigma}+\mathcal{F_{\mu\nu}}g_{\lambda\sigma}-4\delta_{[\mu}^{\alpha}g_{\nu][\lambda}\delta_{\sigma]}^{\beta}\left(\nabla_{\alpha}A_{\beta}+A_{\alpha}A_{\beta}-\frac{A_{\alpha}A^{\alpha}}{2}g_{\alpha\beta}\right)\end{equation}
}{\large \par}

{\large After defining the Weyl covariant Riemann curvature tensor,
it is natural to define the }\emph{\large Weyl-covariantized Ricci
tensor}{\large{} $\mathcal{R}_{\mu\nu}$ and }\emph{\large Weyl-covariantized
Ricci scalar}{\large{} $\mathcal{R}$ by,}{\large \par}

{\large \begin{equation}
\mathcal{R}_{\mu\nu}=R_{\mu\nu}+\left(d-2\right)\left(\nabla_{\mu}A_{\nu}+A_{\mu}A_{\nu}-A_{\alpha}A^{\alpha}g_{\mu\nu}\right)+g_{\mu\nu}\nabla_{\lambda}A^{\lambda}+\mathcal{F_{\mu\nu}}\end{equation}
}{\large \par}

{\large \begin{equation}
\mathcal{R}=R+2\left(d-1\right)\nabla_{\lambda}A^{\lambda}-\left(d-2\right)\left(d-1\right)A_{\lambda}A^{\lambda}\end{equation}
}{\large \par}

{\large Note that we will work with a flat metric on the boundary,
hence, the Riemann curvature tensor, the Ricci tensor, and the Ricci
scalar of the boundary are zero (However, their Weyl definitions are
not zero).}{\large \par}

{\large In order to evaluate fluid dynamics to the second order, we
will need to take all sorts of terms with two Weyl covariant derivatives.
This work was done in \cite{key-15} and can also be found in \cite{key-14}
which we use for our notations.}{\large \par}

{\large \newpage{}}{\large \par}

\chapter{{\LARGE Fluid Gravity Correspondence}}

{\large The fluid gravity correspondence is the AdS/CFT or gauge/gravity
duality in a long wavelength}%
{}{\large{} regime. This duality relates a particular strongly coupled
non-abelian gauge theory in $d$ dimensions to string theory, which
in a certain regime reduces to classical gravity in $d+1$ dimensions.
The regimes we will work with are the planar limit in the field theory
and the long wavelength regime. The latter is a small Knudsen number
regime, and the CFT can be described effectively by a hydrodynamic
description. Hence, we have a duality between a $d$-dimensional relativistic
fluid to a classical gravity in $d+1$ dimensions which is just Einstein's
equations with a negative cosmological constant in an asymptotically
Anti de Sitter space-time. We can describe this duality holographically
and relate to the fluid as {}``living on the boundary'' of the whole
$d+1$ space-time, which we will refer to as the bulk. Moreover, the
small Knudsen number suggests an expansion of the fields, from both
side of the duality, in a derivative expansion. This will allow us
to solve Einstein's equations order by order. It turns out that parts
of Einstein's equations at a certain order implement the stress tensor
conservation equations of the fluid at a lower order. This statement
will be checked in the following chapters, but not before we will
give the basic set up of the bulk space-time geometry in this chapter.
Two reviews on the gauge/gravity duality can be found in \cite{key-9,key-10}.}{\large \par}

\section{{\large Preliminaries: Schwarzschild black holes in $AdS_{d+1}$}}

{\large We would like to look at the dual description of fluid from
the gravity perspective. In order to do so, we will have to solve
Einstein's equations with a negative cosmological constant with a
particular choice of units ($R_{AdS}=1)$ and we get:}{\large \par}

{\large \begin{eqnarray}
E_{ab} & =R_{ab}-\frac{1}{2}Rg_{ab}-\frac{d\left(d-1\right)}{2}g_{ab}=T_{ab}^{matter}=0\nonumber \\
 & \Longrightarrow R_{ab}+dg_{ab}=0,\quad R=-d\left(d+1\right)\label{eq:Einstein's equations}\end{eqnarray}
}{\large \par}

{\large One solution to these equations is just the $AdS_{d+1}$ solution
which is dual to a vacuum state in the CFT. Another class of solutions
is described by the \textquotedbl{}boosted Schwarzschild black branes\textquotedbl{},}{\large \par}

{\large \begin{equation}
ds^{2}=-2u_{\mu}dx^{\mu}dr-r^{2}f\left(br\right)u_{\mu}u_{\nu}dx^{\mu}dx^{\nu}+r^{2}P_{\mu\nu}dx^{\mu}dx^{\nu}\label{eq:static metrics}\end{equation}
with,}{\large \par}

{\large \begin{eqnarray}
 &  & f\left(r\right)=1-\frac{1}{r^{d}},\quad u^{\mu}=\left(\gamma,\gamma u^{i}\right),\quad\gamma=\frac{1}{\sqrt{1-\left(u^{i}\right)^{2}}}\nonumber \\
 &  & b=\frac{d}{4\pi T},\quad\left(u^{i}\right)^{2}=u^{i}u_{i}\end{eqnarray}
}{\large \par}

{\large This solution is dual to a CFT at a finite temperature $T$,
where the velocity $u^{i}$ and the temperature are constants.}{\large \par}

{\large The metrics (\ref{eq:static metrics}) describe the uniform
black brane in an asymptotically Anti de Sitter space-time written
in the ingoing Eddington-Finkelstein coordinates}%
\footnote{These coordinates exhibit a regular solution at the event horizon
for all orders of expansion.%
}{\large , at temperature $T$, moving at velocity $u^{i}$.}{\large \par}

{\large The sets of solutions characterized by $d$ parameters give
us $d$ possible different solutions. The parameters are just the
temperature and the $d$ parameters which define the $d$-velocity,
so we have $d+1$ parameters, and because of the normalization (\ref{eq:normalization})
they are reduced to $d$ parameters. }{\large \par}

\begin{figure}
{\large \includegraphics[scale=0.5]{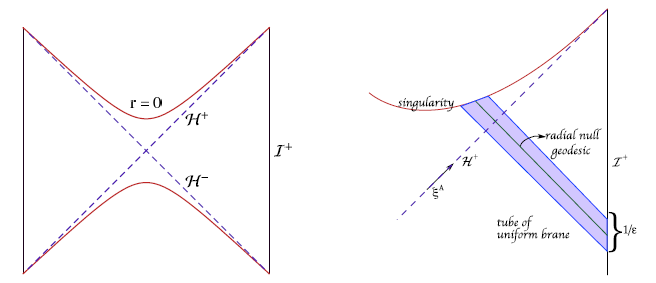}}{\large \par}

{\large \caption{{\large \label{figure 1}}Penrose diagram of the uniform black brane
and the causal structure of the space-times dual to fluid mechanics
illustrating the tube structure. The dashed line in the second figure
denotes the future event horizon, while the shaded tube indicates
the region of space-time over which the solution is well approximated
by a tube of the uniform black brane. (Taken from \cite{key-9}).}
}
\end{figure}

{\large If we had written the solution in Schwarzschild type coordinates}%
\footnote{The unboosted solution in Schwarzschild type coordinates

\[
ds^{2}=-r^{2}f\left(br\right)dt^{2}+\frac{dr^{2}}{r^{2}f\left(br\right)}+r^{2}\delta_{ij}dx^{i}dx^{j}\]
with $\delta_{ij}$ is Kronecker's delta. Note that it is not exhibits
regularity at the event horizon location.%
}{\large{} we could easily have read from the metric the location of
the event horizon $r=\frac{1}{b}$. One can look on the left image
of figure \ref{figure 1} in order to understand in a pictorial way
the casual structure of space-time.}{\large \par}

{\large }%
{\large \par}

\section{{\large Away from global equilibrium}}

{\large Now we would like to consider an out-of-equilibrium black
brane in such manner that the dual description will give us hydrodynamics
of a viscous fluid \cite{key-3}. In order to do so, we promote the
parameters to be dependent on the boundary coordinates $x^{\mu}$,
but not on the radial direction $r$, and we get from (\ref{eq:static metrics}):}{\large \par}

{\large \begin{equation}
ds^{2}=-2u_{\mu}\left(x^{\alpha}\right)dx^{\mu}dr-r^{2}f\left(b\left(x^{\alpha}\right)r\right)u_{\mu}\left(x^{\alpha}\right)u_{\nu}\left(x^{\alpha}\right)dx^{\mu}dx^{\nu}+r^{2}P_{\mu\nu}\left(x^{\alpha}\right)dx^{\mu}dx^{\nu}\label{eq:perturbed  metrics}\end{equation}
}{\large \par}

{\large However (\ref{eq:perturbed  metrics}), denoted as $g^{\left(0\right)}\left(u\left(x^{\alpha}\right),b\left(x^{\alpha}\right)\right)$}%
\footnote{Consider the fields as the zeroth order of \ref{eq:fields expansion}%
}{\large , is not a solution of Einstein's equations.}{\large \par}

{\large The procedure to satisfy Einstein's equations~goes as follow:}{\large \par}
\begin{enumerate}
\item {\large Expand the velocity and temperature fields in a small parameter.}{\large \par}
\item {\large Use the ansatz so that after a while the system relaxes and
go back to the static solution (\ref{eq:static metrics}). }{\large \par}
\item {\large Calculate the metric order-by-order to satisfy Einstein's
equations. }{\large \par}
\end{enumerate}
{\large We will do the expansions in the manner of effective field
theory. As we have done in (\ref{eq:stress tensor expansion}), we
will take all possible derivatives along the boundary coordinates
(i.e., $x^{\mu})$ to get the expansions,}{\large \par}

{\large \begin{equation}
u^{\mu}\left(x^{\alpha}\right)=\sum_{l=0}^{\infty}u_{\left(l\right)}^{\mu}\left(x^{\alpha}\right),\quad b\left(x^{\alpha}\right)=\sum_{l=0}^{\infty}b_{\left(l\right)}\left(x^{\alpha}\right)\label{eq:fields expansion}\end{equation}
}{\large \par}

{\large It is possible to prove that every derivative will come with
a factor of $b$, and if the fields are changing slowly when looking
on a large length scale $L$, we have that every term in the expansions
is proportional  to $\left(\frac{b}{L}\right)^{l}\sim\left(\frac{1}{TL}\right)^{l}$.
In other words, in order that these expansions be justified we will
require the Knudsen number to be small \ref{eq:expansion requirement}.}{\large \par}

\begin{figure}
{\large \includegraphics[scale=0.6]{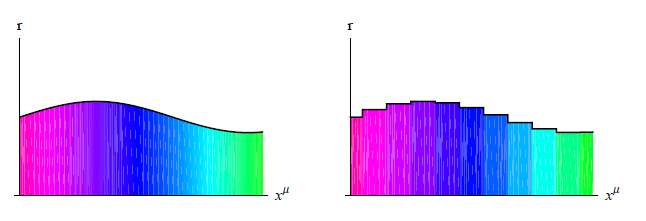}}{\large \par}

{\large \caption{{\large \label{figure 2}}Cartoon of \textquoteleft{}tubewise approximation\textquoteright{}
of slowly-varying configuration by a corresponding piecewise-constant
one. (Taken from \cite{key-10}).}
}
\end{figure}

{\large This regime of expansions can be visualized as tubes that
go along the radial direction and their basis is on the boundary (figure
\ref{figure 1} right picture). In those tubes we can define a definite
velocity and definite temperature (figure \ref{figure 2}).}{\large \par}

{\large }%
{\large \par}

{\large Now we will plug the expansions (\ref{eq:fields expansion})
into the metric and get $g_{ab}=\sum_{l=0}^{\infty}g_{ab}^{\left(l\right)}$.
Then we will require that this metric solves Einstein's equation and
we can determine the metric order-by-order in an iterative form.}{\large \par}

{\large The Einstein's equations can be written in the form:}{\large \par}

{\large \begin{equation}
\mathbb{H\left[\text{\ensuremath{g^{\left(0\right)}\left(b_{\left(0\right)},u_{\left(0\right)}^{\mu}\right)}}\right]}g_{ab}^{\left(n\right)}\left(x^{\alpha}\right)=s_{n}\label{eq:operator form of Einstein's equations}\end{equation}
where $\mathbb{H}$ is a linear second order differential operator
in the $r$ variable alone. Note that it does not depend on $n$,
so that it is the same operator in any order of the expansion. Moreover,
the precise form of this operator at the point $x^{\mu}$ depends
only on the values of $b_{\left(0\right)}$ and $u_{\left(0\right)}^{\mu}$
at $x^{\mu}$ but not on the derivatives of these functions at that
point. The source term $s_{n}$ is a regular source term which contains
complicated combinations of derivatives of the fields and the fields
themselves and it is different in any order of expansion.}{\large \par}

{\large It turns out that it is possible to classify Einstein's equations
into two categories:}{\large \par}
\begin{enumerate}
\item {\large the }\emph{\large constraint equations}{\large{} $E_{\mu}^{r}=0$.}{\large \par}
\item {\large the }\emph{\large dynamical equations}{\large{} $E_{r}^{r}=0,\: E_{\mu}^{\nu}=0$.}{\large \par}
\end{enumerate}
{\large The $d$ constraint equations are defined as the equations
that are of first order in $r$ derivatives. It follows that they
are connecting the derivatives of the temperature field with the derivatives
of the velocity field. Hence we can use them in order to eliminate
the temperature derivatives in all order of expansion. The remaining
$\frac{d\left(d+1\right)}{2}$ dynamical equations are determined
by the next order metric}%
\footnote{There is a redundancy among the remaining equations which leaves $\frac{d\left(d-1\right)}{2}$
independent \textquoteleft{}dynamical\textquoteright{} equations,
this freedom is eliminating by choosing the following gauge condition:

\[
g_{rr}=0,\quad g_{r\mu}=-u_{\mu}.\]
}{\large . Then all together we have $\frac{\left(d+1\right)\left(d+2\right)}{2}$
which is the number of equations that (\ref{eq:operator form of Einstein's equations})
possess. }{\large \par}

\section{{\large The Bulk Metric}}

{\large In this section we will present the general form of metrics
that can describe the bulk geometry of AdS and are also in accord
with the conformal fluid on the boundary of the AdS space-time. Then
we will present the second order solution of the metric which satisfy
(\ref{eq:Einstein's equations}) in a Weyl covariant formulation \cite{key-1}.}{\large \par}

{\large We start by writing the most general metric that satisfies
the gauge choice of the previous section:}{\large \par}

{\large \begin{equation}
ds^{2}=-2u_{\mu}\left(x^{\alpha}\right)dx^{\mu}\left(dr+\mathcal{V}_{\nu}\left(r,x^{\alpha}\right)dx^{\nu}\right)+\mathfrak{G_{\mu\nu}}\left(r,x^{\alpha}\right)dx^{\mu}dx^{\nu},\label{eq:general metric}\end{equation}
where $\mathfrak{G_{\mu\nu}}$ is transverse to the fluid velocity,
i.e., $u^{\mu}\mathfrak{G_{\mu\nu}}=0$.}{\large \par}

{\large We will also present the inverse of the bulk metric,}{\large \par}

{\large \begin{eqnarray}
u^{\mu}\left[\left(\partial_{\mu}-\mathcal{V}_{\mu}\partial_{r}\right)\otimes\partial_{r}+\partial_{r}\otimes\left(\partial_{\mu}-\mathcal{V}_{\mu}\partial_{r}\right)\right]\nonumber \\
+\left(\mathfrak{G}^{-1}\right)^{\mu\nu}\left(\partial_{\mu}-\mathcal{V}_{\mu}\partial_{r}\right)\otimes\left(\partial_{\nu}-\mathcal{V}_{\nu}\partial_{r}\right)\end{eqnarray}
or in another form,}{\large \par}

{\large \begin{equation}
ds^{2}=2\left(u^{\mu}-\left(\mathfrak{G}^{-1}\right)^{\mu\nu}\mathcal{V}_{\nu}\right)\partial_{\mu}\otimes\partial_{r}+\left(\left(\mathfrak{G}^{-1}\right)^{\mu\nu}\mathcal{V}_{\mu}\mathcal{V}_{\nu}-2u^{\mu}\mathcal{V}_{\mu}\right)\partial_{r}\otimes\partial_{r}+\left(\vartheta^{-1}\right)^{\mu\nu}\partial_{\mu}\otimes\partial_{\nu}\end{equation}
where the symmetric tensor $\left(\mathfrak{G}^{-1}\right)^{\mu\nu}$
is uniquely defined by the relations: $u_{\mu}\left(\mathfrak{G}^{-1}\right)^{\mu\nu}=0$
and $\left(\mathfrak{G}^{-1}\right)^{\mu\lambda}\mathfrak{G_{\lambda\nu}}=P_{\nu}^{\mu}$.}{\large \par}

{\large We will now write the solution to the second order,}{\large \par}

{\large \begin{eqnarray*}
\mathcal{V}_{\mu} & = & rA_{\mu}+\frac{1}{d-2}\left[D_{\lambda}\omega_{\:\mu}^{\lambda}-D_{\lambda}\sigma_{\:\mu}^{\lambda}+\frac{\mathcal{R}}{2\left(d-1\right)}u_{\mu}\right]-\frac{2L\left(br\right)}{\left(br\right)^{d-2}}P_{\mu}^{\nu}D_{\lambda}\sigma_{\:\nu}^{\lambda}\\
 &  & -\frac{u_{\mu}}{2\left(br\right)^{d}}\left[r^{2}\left(1-\left(br\right)^{d}\right)-\frac{1}{2}\omega_{\alpha\beta}\omega^{\alpha\beta}-\left(br\right)^{2}K_{2}\left(br\right)\frac{\sigma_{\alpha\beta}\sigma^{\alpha\beta}}{d-1}\right]\end{eqnarray*}
}{\large \par}

{\large \begin{eqnarray*}
\mathfrak{G_{\mu\nu}} & = & r^{2}P_{\mu\nu}-\omega_{\mu}^{\:\lambda}\omega_{\lambda\nu}\\
 &  & +2\left(br\right)^{2}F\left(br\right)\left[\frac{1}{b}\sigma_{\mu\nu}+F\left(br\right)\sigma_{\mu}^{\:\lambda}\sigma_{\lambda\nu}\right]-2\left(br\right)^{2}K_{1}\left(br\right)\frac{\sigma_{\alpha\beta}\sigma^{\alpha\beta}}{d-1}P_{\mu\nu}\\
 &  & -2\left(br\right)^{2}H_{1}\left(br\right)\left[u^{\lambda}D_{\lambda}\sigma_{\mu\nu}+\sigma_{\mu}^{\:\lambda}\sigma_{\lambda\nu}-\frac{\sigma_{\alpha\beta}\sigma^{\alpha\beta}}{d-1}P_{\mu\nu}\right]\\
 &  & +2\left(br\right)^{2}H_{2}\left(br\right)\left[u^{\lambda}D_{\lambda}\sigma_{\mu\nu}+\omega_{\mu}^{\:\lambda}\sigma_{\lambda\nu}+\omega_{\nu}^{\:\lambda}\sigma_{\mu\lambda}\right]\end{eqnarray*}
and,}{\large \par}

{\large \begin{eqnarray*}
\left(\mathfrak{G}^{-1}\right)^{\mu\nu} & = & \frac{1}{r^{2}}P^{\mu\nu}+\frac{1}{r^{4}}\omega^{\mu\lambda}\omega_{\lambda}^{\:\nu}\\
 &  & -\frac{2b^{2}}{r^{2}}F\left(br\right)\left[\frac{1}{b}\sigma^{\mu\nu}-F\left(br\right)\sigma_{\:\lambda}^{\mu}\sigma^{\lambda\nu}\right]+\frac{2b^{2}}{r^{2}}K_{1}\left(br\right)\frac{\sigma_{\alpha\beta}\sigma^{\alpha\beta}}{d-1}P^{\mu\nu}\\
 &  & +\frac{2b^{2}}{r^{2}}H_{1}\left(br\right)\left[u^{\lambda}D_{\lambda}\sigma^{\mu\nu}+\sigma_{\:\lambda}^{\mu}\sigma^{\lambda\nu}-\frac{\sigma_{\alpha\beta}\sigma^{\alpha\beta}}{d-1}P^{\mu\nu}\right]\\
 &  & -\frac{2b^{2}}{r^{2}}H_{2}\left(br\right)\left[u^{\lambda}D_{\lambda}\sigma^{\mu\nu}+\omega_{\:\lambda}^{\mu}\sigma^{\lambda\nu}+\omega_{\:\lambda}^{\nu}\sigma^{\mu\lambda}\right]\end{eqnarray*}
with,}{\large \par}

{\large \[
F\left(br\right)\equiv\int_{br}^{\infty}\frac{y^{d-1}-1}{y\left(y^{d}-1\right)}dy\]
}{\large \par}

{\large \[
H_{1}\left(br\right)\equiv\int_{br}^{\infty}\frac{y^{d-2}-1}{y\left(y^{d}-1\right)}dy\]
}{\large \par}

{\large \[
H_{2}\left(br\right)\equiv\frac{1}{2}F\left(br\right)^{2}-\int_{br}^{\infty}\frac{d\xi}{\xi\left(\xi^{d}-1\right)}\int_{1}^{\xi}\frac{y^{d-2}-1}{y\left(y^{d}-1\right)}dy\]
}{\large \par}

{\large \[
K_{1}\left(br\right)\equiv\int_{br}^{\infty}\frac{d\xi}{\xi^{2}}\int_{\xi}^{\infty}dy\, y^{2}F^{\prime}\left(y\right)^{2}\]
}{\large \par}

{\large \begin{eqnarray*}
K_{2}\left(br\right) & \equiv & \int_{br}^{\infty}\frac{d\xi}{\xi^{2}}\left[\vphantom{\int_{\xi}^{\infty}}1-\xi\left(\xi-1\right)F^{\prime}\left(\xi\right)-2\left(d-1\right)\xi^{d-1}\right.\\
 &  & \left.+\left(2\left(d-1\right)\xi^{d}-\left(d-2\right)\right)\int_{\xi}^{\infty}dy\, y^{2}F^{\prime}\left(y\right)^{2}\right]\end{eqnarray*}
}{\large \par}

{\large \[
L\left(br\right)\equiv\int_{br}^{\infty}\xi^{d-1}d\xi\int_{\xi}^{\infty}dy\,\frac{y-1}{y^{3}\left(y^{d}-1\right)}\]
}{\large \par}

{\large \newpage{}}{\large \par}

\chapter{{\LARGE Horizon Dynamics}}

{\large Given a manifold of space-time one would like to investigate
the geodesic dynamics equations of that manifold. To do that one has
to solve the Einstein equations. However, it is possible to consider
slices of that manifold, for instance hyper-surfaces. Einstein's equations
projected onto a hyper-surface take a simpler form called the Gauss-Codazzi
equations.}{\large \par}

{\large In order to understand the geodesic dynamics on hyper-surfaces
of the space-time geometry (Gauss-Codazzi equations), we need to introduce
some basic quantities that characterize the hyper-surfaces \cite{key-13}
(induced metric, extrinsic curvature, etc.).}{\large \par}

\section{{\large Geometric Preliminaries}}

{\large A }\emph{\large hyper-surface}{\large{} is a sub-manifold that
can either be space-like, time-like, or null}%
\footnote{For the null case refer to \cite{key-11,key-12}%
}{\large{} (we will almost not deal with the latter in this section).
A particular hyper-surface $\mathcal{S}$ is selected either by imposing
a restriction on the coordinates, $\Phi\left(x^{a}\right)=0$, or
by providing a parametric equation of the form $x^{a}=x^{a}\left(y^{\mu}\right)$.}{\large \par}

{\large The normal vector to the surface $\mathcal{S}$ is defined
by $\Phi_{,a}$, because the value of $\Phi$ changes only in the
direction orthogonal to $\mathcal{S}$.}{\large \par}

{\large A }\emph{\large unit normal}{\large{} $\hat{m}^{a}$ is defined
by:}{\large \par}

{\large \begin{equation}
\hat{m}_{a}=\frac{\Phi_{,a}}{\sqrt{\left|g^{ab}\Phi_{,a}\Phi_{,b}\right|}}\label{eq:normal vector}\end{equation}
where we demand that $\hat{m}^{a}$ point in the direction of increasing
$\Phi$: $\hat{m}^{a}\Phi_{,a}>0$, and we get its norm to be:}{\large \par}

{\large \begin{equation}
\hat{m}^{a}\hat{m}_{a}=\begin{cases}
-1 & if\: S\: is\: space-like\\
+1 & if\: S\: is\: time-like\end{cases}\end{equation}
}{\large \par}

{\large In the case of a null surface we can only define a class of
null normal co-vectors, in the following way:}{\large \par}

{\large \begin{equation}
\ell_{a}=f\,\Phi_{,a}\label{eq:null vector defintion}\end{equation}
where $f$ is an arbitrary function. The norm of the class of vectors
is:}{\large \par}

{\large \begin{equation}
\ell^{a}\ell_{a}=0\end{equation}
.}{\large \par}

{\large Now we would like to define a metric that would be intrinsic
to the hyper-surface. The }\emph{\large induced metric}{\large{} $h_{\mu\nu}$
to a hyper-surface $\mathcal{S}$ is obtained by restricting the line
element to displacements confined to the hyper-surface. Using the
parametric equations $x^{a}=x^{a}\left(y^{\mu}\right)$, we find that
the vectors,}{\large \par}

{\large \begin{equation}
e_{\mu}^{a}=\frac{\partial x^{a}}{\partial y^{\mu}}\label{eq:basis vectors}\end{equation}
are tangent to curves contained in $\mathcal{S}$. (This means that
$e_{\mu}^{a}\hat{m}_{a}=0$}%
\footnote{$e_{\mu}^{a}\hat{m}_{a}=0\propto\frac{\partial x^{a}}{\partial y^{\mu}}\frac{\partial\Phi}{\partial x^{a}}=\frac{\partial\Phi}{\partial y^{\mu}}=0$.
It is zero because the function is constant in the directions of the
surface.%
}{\large ). Therefore, we now write the restriction of the line element
for displacements within $\mathcal{S}$:}{\large \par}

{\large \begin{eqnarray}
ds^{2} & = & g_{ab}dx^{a}dx^{b}\nonumber \\
 & = & g_{ab}\left(\frac{\partial x^{a}}{\partial y^{\mu}}dy^{\mu}\right)\left(\frac{\partial x^{b}}{\partial y^{\nu}}dy^{\nu}\right)\\
 & = & h_{\mu\nu}dy^{\mu}dy^{\nu}\nonumber \end{eqnarray}
where}{\large \par}

{\large \begin{equation}
h_{\mu\nu}=g_{ab}e_{\mu}^{a}e_{\nu}^{b}\end{equation}
}{\large \par}

{\large Note that the induced metric is a scalar under bulk space-time
coordinate transformations, $x^{a}\rightarrow x^{a'}$. However, it
transforms as a tensor under the hyper-surface coordinate transformations,
$y^{\mu}\rightarrow y^{\mu'}$. }{\large \par}

{\large With the aid of the induced metric we can write the decomposition
of the bulk metric:}{\large \par}

{\large \begin{equation}
g^{ab}=h^{ab}\pm\hat{m}^{a}\hat{m}^{b}\label{eq:decomposition}\end{equation}
with:}{\large \par}

{\large \begin{equation}
h^{ab}=h^{\alpha\beta}e_{\alpha}^{a}e_{\beta}^{b}\label{eq:upper induced metric}\end{equation}
where the plus sign in (\ref{eq:decomposition}) represents the time-like
hyper-surfaces, and the minus represents the space-like hyper-surfaces.}{\large \par}

{\large This decomposition can be defined to every hyper-surface that
is not null.}{\large \par}

{\large We now would like to understand how tangent tensor fields
are differentiated. First we define such fields:}{\large \par}

{\large A}\emph{\large{} tangent tensor field}{\large{} $A^{ab\cdots}$
is a tensor field that is defined only on $\mathcal{S}$ and is purely
tangent to the hyper-surface. Such fields admit the following decomposition:}{\large \par}

{\large \begin{equation}
A^{ab\cdots}=A^{\mu\nu\cdots}e_{\mu}^{a}e_{\nu}^{b}\cdots\label{eq:tangent tensor}\end{equation}
}{\large \par}

{\large Equation (\ref{eq:tangent tensor}) implies that $A^{ab\cdots}\hat{m}_{a}=A^{ab\cdots}\hat{m}_{b}=\cdots=0$
which confirm that $A^{ab\cdot\cdot\cdot}$ is tangent to the hyper-surface.
The tangent tensor $A^{\mu\nu}$ indices are lowered and raised with
the induced metric on the hyper-surface $h_{\mu\nu}$ and $h^{\mu\nu}$.}{\large \par}

{\large Now we can define an }\emph{\large intrinsic covariant derivative}{\large .
We will use a tangent vector field for simplicity: $A^{a}=A^{\mu}e_{\mu}^{a},\quad A^{a}\hat{m}_{a}=0,\quad A_{\mu}=A_{a}e_{\mu}^{a}$.
The intrinsic covariant derivative of $A_{\mu}$ is defined by the
projection of $A_{a;b}$ onto the hyper-surface:}{\large \par}

{\large \begin{equation}
A_{\mu|\nu}\equiv A_{a;b}e_{\mu}^{a}e_{\nu}^{b}\label{eq:definition of intrinsic covariant derivative}\end{equation}
}{\large \par}

{\large It is possible to prove that equation (\ref{eq:definition of intrinsic covariant derivative})
can take the form of,}{\large \par}

{\large \begin{equation}
A_{\mu|\nu}\equiv A_{\mu,\nu}-\Gamma_{\mu\nu}^{\sigma}A_{\sigma}\end{equation}
where }{\large \par}

{\large \begin{equation}
\Gamma_{\mu\nu}^{\sigma}=h^{\sigma\lambda}\left(h_{\mu\sigma,\nu}+h_{\nu\sigma,\mu}-h_{\mu\nu.\sigma}\right)\end{equation}
}{\large \par}

{\large We can now introduce the }\emph{\large extrinsic curvature}{\large{}
of a hyper-surface, which is just the projection of the covariant
derivative of the normal to the hyper-surface onto the hyper-surface
$S$:}{\large \par}

{\large \begin{equation}
K_{\mu\nu}\equiv\hat{m}_{a;b}e_{\mu}^{a}e_{\nu}^{b}\end{equation}
}{\large \par}

{\large This quantity is a symmetric tensor and it tells us how the
hyper-surface $S$ is embedded in the bulk. We can see this by taking
its trace: $K=h^{\mu\nu}K_{\mu\nu}=\hat{m}_{;a}^{a}$ (we used the
decomposition equation (\ref{eq:decomposition})) which is the expansion}%
\footnote{This term will be explained in chapter 6.%
}{\large{} of a congruence of geodesics that intersect the hyper-surface
$S$ orthogonally. When $K>0$ (the congruence is diverging), the
hyper-surface is }\emph{\large convex}{\large{} and if $K<0$ (the
congruence is converging), the hyper-surface is }\emph{\large concave. }{\large \par}

\section{{\large Gauss-Codazzi equations}}

{\large We are ready to introduce the Einstein equations on a hyper-surface
$S$. We start by defining a purely intrinsic curvature tensor by
the relation:}{\large \par}

{\large \begin{equation}
A_{|\mu\nu}^{\sigma}-A_{|\nu\mu}^{\sigma}=R_{\lambda\nu\mu}^{\sigma}A^{\lambda}\end{equation}
where the explicit expression for the purely intrinsic curvature tensor
is:}{\large \par}

{\large \begin{equation}
R_{\lambda\nu\mu}^{\sigma}=\Gamma_{\lambda\mu,\nu}^{\sigma}-\Gamma_{\lambda\nu,\mu}^{\sigma}+\Gamma_{\alpha\nu}^{\sigma}\Gamma_{\lambda\mu}^{\alpha}-\Gamma_{\alpha\mu}^{\sigma}\Gamma_{\lambda\nu}^{\alpha}\end{equation}
}{\large \par}

{\large We now want to establish a connection between the intrinsic
curvature tensor to the Riemann curvature tensor defined in the bulk
geometry of space-time. It can be proven \cite{key-13} that the relation
is:}{\large \par}

{\large \begin{equation}
R_{abc}^{d}e_{\alpha}^{a}e_{\beta}^{b}e_{\gamma}^{c}=R_{\alpha\beta\gamma}^{\sigma}e_{\sigma}^{d}\pm\left(K_{\alpha\beta|\gamma}-K_{\alpha\gamma|\beta}\right)\hat{m}^{d}\pm K_{\alpha\beta}\hat{m}_{;c}^{d}e_{\gamma}^{c}\mp K_{\alpha\gamma}\hat{m}_{;b}^{d}e_{\beta}^{b}\label{eq:gauss-codazzi not contracted}\end{equation}
where again the plus sign represents the time-like hyper-surfaces
and the minus represents the space-like hyper-surfaces.}{\large \par}

{\large Projecting along $m_{d}$ and using the symmetry of Riemann
curvature tensor}%
\footnote{$R_{abcd}=-R_{bacd}=-R_{abdc}=R_{cdab}$%
}{\large{} gives:}{\large \par}

{\large \begin{equation}
R_{adcb}\hat{m}^{d}e_{\alpha}^{a}e_{\beta}^{b}e_{\gamma}^{c}=\left(K_{\alpha\beta|\gamma}-K_{\alpha\gamma|\beta}\right)\label{eq:guess-codazzi projected and not contracted}\end{equation}
}{\large \par}

{\large We will now contract the indices by using the metric }\textbf{\large $h^{\alpha\gamma}$}{\large{}
with equation (\ref{eq:guess-codazzi projected and not contracted}).
Then, by using equation (\ref{eq:decomposition}) and the symmetry
of Riemann curvature tensor, we get:}{\large \par}

{\large \begin{equation}
R_{ab}\hat{m}^{a}e_{\beta}^{b}=\left(K_{\beta|\alpha}^{\alpha}-K_{,\beta}\right)\label{eq:guess-codazzi contracted after getting  rid of the metric}\end{equation}
}{\large \par}

{\large If we look at the bulk metric $g_{ab}$ contracted with $\hat{m}^{a}e_{\beta}^{b}$
we get,}{\large \par}

{\large \begin{equation}
g_{ab}\hat{m}^{a}e_{\beta}^{b}=\hat{m}_{b}e_{\beta}^{b}=0\label{eq:getting rid of the metric}\end{equation}
}{\large \par}

{\large This means that we can replace the left hand side of (\ref{eq:guess-codazzi contracted after getting  rid of the metric})
with $\left(R_{ab}+Cg_{ab}\right)\hat{m}^{a}e_{\beta}^{b}$, where
$C$ is an arbitrary coefficient, because every term that is proportional
to the metric will vanish. For instance, we can replace the left hand
side with the Einstein tensor $G_{ab}=R_{ab}-\frac{1}{2}Rg_{ab}$
to get the }\emph{\large Gauss-Codazzi equations,}{\large \par}

{\large \begin{equation}
G_{ab}\hat{m}^{a}e_{\beta}^{b}=K_{\beta|\alpha}^{\alpha}-K_{,\beta}\label{eq:guess-codazzi contracted}\end{equation}
}{\large \par}

{\large Because we are solving the Einstein equations with a cosmological
constant we would like to replace the left hand side of (\ref{eq:guess-codazzi contracted after getting  rid of the metric})
with $E_{ab}$, which we introduced in equation (\ref{eq:Einstein's equations}),
and we get,}{\large \par}

{\large \begin{equation}
E_{ab}\hat{m}^{a}e_{\beta}^{b}=K_{\beta|\alpha}^{\alpha}-K_{,\beta}\label{eq:guess-codazzi contracted for AdS}\end{equation}
}{\large \par}

{\large As mentioned in the equations (\ref{eq:guess-codazzi contracted})
and (\ref{eq:guess-codazzi contracted for AdS}) we are eventually
reducing these equations, because of (\ref{eq:getting rid of the metric}),
to (\ref{eq:guess-codazzi contracted after getting  rid of the metric})
and these are the equations we have to solve in order to find the
geodesic dynamic equations.}{\large \par}

\section{{\large The Constraint equations}}

{\large In the previous chapter we stated and showed that Einstein's
equations can be divided into two categories, the dynamical equations
and the constraint equations. In this section we will deal with the
latter.}{\large \par}

\underbar{\large Claim}{\large : The constraint equations $E_{\mu}^{r}=0$
with the metric (\ref{eq:general metric}), which solves the dynamic
Einstein's equation, can be written in the form:}{\large \par}

{\large \begin{equation}
E_{ab}\xi^{b}e_{\mu}^{a}=0\label{eq:Claim}\end{equation}
where $\xi^{b}$ is normal to any hyper-surface $S$ (time-like, space-like
and null) that is defined by $\Phi=r-r\left(x^{\alpha}\right)=0$.
Thus $\xi^{b}$ is defined by $\xi^{b}=f\, g^{ba}\Phi_{,a}$, where
$f$ is an arbitrary normalization function.}{\large \par}

\underbar{\large Proof}{\large : Let us set $f=1$ for convenience
reasons and look at the LHS of (\ref{eq:Claim}):}{\large \par}

{\large \begin{eqnarray*}
E_{ab}\xi^{b}e_{\mu}^{a} & = & R_{ab}\xi^{b}e_{\mu}^{a}\\
 & = & e_{\mu}^{a}R_{ab}g^{bc}\partial_{c}\Phi\\
 & = & e_{\mu}^{a}R_{a}^{r}-e_{\mu}^{a}R_{a}^{\nu}\partial_{\nu}r\left(x^{\alpha}\right)\\
 & = & R_{\mu}^{r}+R_{r}^{r}\partial_{\mu}r\left(x^{\alpha}\right)-R_{\mu}^{\nu}\partial_{\nu}r\left(x^{\alpha}\right)-R_{r}^{\nu}\partial_{\mu}r\left(x^{\alpha}\right)\partial_{\nu}r\left(x^{\alpha}\right)\\
 & = & R_{\mu}^{r}-d\partial_{\mu}r\left(x^{\alpha}\right)+d\delta_{\mu}^{\nu}\partial_{\nu}r\left(x^{\alpha}\right)-\left(g^{\nu\sigma}R_{\sigma r}+g^{\nu r}R_{rr}\right)\partial_{\mu}r\left(x^{\alpha}\right)\partial_{\nu}r\left(x^{\alpha}\right)\\
 & = & R_{\mu}^{r}+dg^{\nu\sigma}g_{\sigma r}\partial_{\mu}r\left(x^{\alpha}\right)\partial_{\nu}r\left(x^{\alpha}\right)\\
 & = & R_{\mu}^{r}-dg^{\nu\sigma}u_{\sigma}\partial_{\mu}r\left(x^{\alpha}\right)\partial_{\nu}r\left(x^{\alpha}\right)\\
 & = & R_{\mu}^{r}\end{eqnarray*}
where we got the fifth line by using Einstein's equations (\ref{eq:Einstein's equations})
and we got the sixth line by using the relation: $g^{\mu\nu}g_{r\nu}=\mathfrak{-G^{\mu\nu}}u_{\nu}=0$.
Using the relation: $E_{\mu}^{r}=R_{\mu}^{r}$ completes the proof.}{\large \par}

\begin{flushright}
{\large $\blacksquare$}
\par\end{flushright}{\large \par}

{\large For the non-null case we can take $\xi^{a}=\hat{m}^{a}$}%
\footnote{Note that the normalization function $f$ can be different from one.%
}{\large{} in (\ref{eq:Claim}) and we get from (\ref{eq:guess-codazzi contracted for AdS})
the Gauss-Codazzi equations with a vanishing matter stress tensor:}{\large \par}

{\large \begin{equation}
K_{\mu|\nu}^{\nu}-K_{,\mu}=0\label{eq:Gauss-Codazzi equations with vanishing matter T_{ab}}\end{equation}
}{\large \par}

{\large If we look at the constraint equations on the boundary with
the normal vector to the boundary surface $\xi^{a}=g^{ar},$ then
the constraint equations, according to \cite{key-4}, take the simple
form of the equations of motion for the boundary fluid (\ref{eq:the equations of motion}).}{\large \par}

{\large The constraint equations just connect the derivatives of the
fluid $d$-velocity with the derivatives of the temperature (or the
inverse temperature $b$), and of course the $d$-velocity and the
temperature fields depend, by our definition (\ref{eq:fields expansion}),
on the coordinates perpendicular to the $r$ coordinate, which are
the boundary coordinates $x^{\alpha}$. For this reason we can claim
that the constraint equations are the same for every hyper-surface
$\Phi=r-r\left(x^{\alpha}\right)$ of space-time. In the following
chapters we check this claim explicitly.}{\large \par}

{\large \newpage{}}{\large \par}

\chapter{{\LARGE The Second Order Constraint Equations}}

\section{Hyper-Surfaces At Constant Radial Location}

{\large In this section we will check the constraint equations to
the first order in derivative expansion for a constant radial coordinate
hyper-surface, i.e., $\Phi=r-R$, where $R$ is constant.}{\large \par}

{\large For reasons of convenience, we will write here only the zeroth
and first order parts of the metric (\ref{eq:general metric})}{\large \par}

{\large \begin{eqnarray}
ds^{2} & = & -2u_{\mu}dx^{\mu}dr-r^{2}f\left(br\right)u_{\mu}u_{\nu}dx^{\mu}dx^{\nu}+r^{2}P_{\mu\nu}dx^{\mu}dx^{\nu}\label{eq:first order metric}\\
 &  & +2r^{2}bF\left(br\right)\sigma_{\mu\nu}dx^{\mu}dx^{\nu}+\frac{2}{d-1}ru_{\mu}u_{\nu}\partial_{\lambda}u^{\lambda}dx^{\mu}dx^{\nu}-ru^{\lambda}\partial_{\lambda}\left(u_{\mu}u_{\nu}\right)dx^{\mu}dx^{\nu}\nonumber \end{eqnarray}
}{\large \par}

{\large The components of the inverse metric to the first order are:}{\large \par}

{\large \begin{equation}
g^{rr}=r^{2}f\left(br\right)-\frac{2}{d-1}r\partial_{\lambda}u^{\lambda},\; g^{r\mu}=u^{\mu}-\frac{1}{r}u^{\lambda}\partial_{\lambda}u^{\mu},\; g^{\mu\nu}=\frac{1}{r^{2}}P^{\mu\nu}-\frac{2bF\left(br\right)}{r^{2}}\sigma^{\mu\nu}\end{equation}
}{\large \par}

{\large Our basis vectors (\ref{eq:basis vectors}) are:}{\large \par}

{\large \begin{equation}
e_{\mu}^{r}=\frac{\partial R}{\partial x^{\mu}}=0,\quad e_{\mu}^{\nu}=\frac{\partial x^{\nu}}{\partial x^{\mu}}=\delta_{\mu}^{\nu}\label{eq:basis for constant R hyper-surface}\end{equation}
}{\large \par}

{\large The normal vector to this surface is:}{\large \par}

{\large \begin{equation}
\hat{m}^{a}=\frac{g^{ab}\partial_{b}\Phi}{\sqrt{\left|g^{ab}\partial_{a}\Phi\partial_{b}\Phi\right|}}=\frac{1}{\sqrt{\left|g^{rr}\right|}}g^{ar}\end{equation}
}{\large \par}

{\large The components of the one-form dual to the normal vector are:}{\large \par}

{\large \begin{equation}
\hat{m}_{r}=\frac{1}{\sqrt{\left|g^{rr}\right|}},\quad\hat{m}_{\mu}=0\end{equation}
}{\large \par}

{\large In order that the last expression will be defined, we demand
$g^{rr}\neq0$. The surface $\Phi$ is time-like if $g^{rr}>0$, or
space-like if $g^{rr}<0$.}{\large \par}

{\large Our goal is to write the constraint equations by using (\ref{eq:Gauss-Codazzi equations with vanishing matter T_{ab}}),
so we will need to find the extrinsic curvature of the hyper-surface.
We will calculate the mixed upper and lower indices of the extrinsic
curvature in the following way:}{\large \par}

{\large \begin{eqnarray}
K_{\nu}^{\mu} & = & h^{\mu\sigma}K_{\sigma\nu}\nonumber \\
 & = & h^{\mu\sigma}\hat{m}_{a;b}e_{\sigma}^{a}e_{\nu}^{b}\nonumber \\
 & = & h^{\mu a}\hat{m}_{a;b}e_{\nu}^{b}\nonumber \\
 & = & \left(g^{\mu a}\mp\hat{m}^{\mu}\hat{m}^{a}\right)\hat{m}_{a;b}e_{\nu}^{b}\\
 & = & g^{\mu a}\hat{m}_{a;b}e_{\nu}^{b}\mp\hat{m}^{\mu}\hat{m}^{a}\hat{m}_{a;b}e_{\nu}^{b}\nonumber \\
 & = & \hat{m}_{;b}^{\mu}e_{\nu}^{b}\nonumber \\
 & = & \nabla_{\nu}\hat{m}^{\mu}\nonumber \end{eqnarray}
where we used (\ref{eq:basis for constant R hyper-surface}) to get
the fourth line and $\hat{m}^{a}\hat{m}_{a;b}=0$ to get the sixth
line. After applying the covariant derivative, we get:}{\large \par}

{\large \begin{equation}
K_{\nu}^{\mu}=\nabla_{\nu}\hat{m}^{\mu}=\partial_{\nu}\hat{m}^{\mu}+\Gamma_{\nu a}^{\mu}\hat{m}^{a}=\partial_{\nu}\hat{m}^{\mu}+\Gamma_{\nu r}^{\mu}\hat{m}^{r}+\Gamma_{\nu\sigma}^{\mu}\hat{m}^{\sigma}\end{equation}
}{\large \par}

{\large In order to compute the constraint equations up to the second
order in derivative expansion, we have to find the constraint equations
that come from the zeroth order metric, and then apply them to the
calculation of the constraint equations that comes from the first
order metric. We show here the explicit form of the extrinsic curvature
to the zeroth order,\begin{equation}
K_{\nu}^{\mu\left(0\right)}=\cdots=\frac{1}{\sqrt{\left|f\left(br\right)\right|}}\left(-\frac{1}{2}\frac{d}{\left(br\right)^{d}}u^{\mu}u_{\nu}+f\left(br\right)\eta_{\nu}^{\mu}\right)\label{eq:zeroth order extrinsic curvature}\end{equation}
and its trace is,}{\large \par}

{\large \begin{equation}
K^{\left(0\right)}=\frac{1}{\sqrt{\left|f\left(br\right)\right|}}\left(\frac{1}{2}\frac{d}{\left(br\right)^{d}}+df\left(br\right)\right)\label{eq:zeroth order trace of extrinsic curvature}\end{equation}
}{\large \par}

{\large Inserting (\ref{eq:zeroth order extrinsic curvature}) and
(\ref{eq:zeroth order trace of extrinsic curvature}) to the Gauss-Codazzi
equations (\ref{eq:Gauss-Codazzi equations with vanishing matter T_{ab}})
we get:}{\large \par}

{\large \begin{eqnarray}
K_{\nu|\mu}^{\mu\left(0\right)}-K_{,\nu}^{\left(0\right)} & = & \frac{d}{2\left(bR\right)^{d}\sqrt{\left|f\left(bR\right)\right|}}\left(2du_{\nu}u^{\mu}\partial_{\mu}\ln b-u^{\mu}\partial_{\mu}u_{\nu}\right.\nonumber \\
 &  & \qquad\qquad\qquad\qquad-u_{\nu}\partial_{\mu}u^{\mu}+\left(d+1\right)\partial_{\nu}\ln b\nonumber \\
 &  & \left.\qquad\qquad\qquad\qquad-dP_{\nu}^{\mu}\partial_{\mu}\ln b\right)=0\label{eq:Gauss-Codazzi R=00003Dconst 1 order}\end{eqnarray}
}{\large \par}

{\large Projecting (\ref{eq:Gauss-Codazzi R=00003Dconst 1 order})
along $u^{\nu}$ we get:}{\large \par}

{\large \begin{eqnarray}
 &  & \frac{d}{2\left(bR\right)^{d}\sqrt{\left|f\left(bR\right)\right|}}\left(\partial_{\mu}u^{\mu}-\left(d-1\right)u^{\nu}\partial_{\nu}\ln b\right)=0\Rightarrow\nonumber \\
 &  & \Rightarrow\partial_{\mu}u^{\mu}-\left(d-1\right)u^{\nu}\partial_{\nu}\ln b=0\label{eq:constraint equation zeroth order 1}\end{eqnarray}
}{\large \par}

{\large Projecting (\ref{eq:Gauss-Codazzi R=00003Dconst 1 order})
along $P_{\sigma}^{\nu}$ we get:}{\large \par}

{\large \begin{eqnarray}
 &  & -\frac{d}{2\left(bR\right)^{d}\sqrt{\left|f\left(bR\right)\right|}}\left(u^{\mu}\partial_{\mu}u_{\sigma}-P_{\sigma}^{\nu}\partial_{\nu}\ln b\right)=0\Rightarrow\nonumber \\
 &  & \Rightarrow u^{\mu}\partial_{\mu}u_{\sigma}-P_{\sigma}^{\nu}\partial_{\nu}\ln b=0\label{eq:constraint equation zeroth order 2}\end{eqnarray}
}{\large \par}

{\large Now we compute the extrinsic curvature to the first order
from the first order bulk metric (\ref{eq:first order metric}),\begin{eqnarray}
K_{\nu}^{\mu\left(0+1\right)} & = & \cdots\nonumber \\
 & = & \frac{1}{r\sqrt{\left|f\left(br\right)\right|}}\left(-\frac{1}{2}r\frac{d}{\left(br\right)^{d}}u^{\mu}u_{\nu}+rf\left(br\right)\eta_{\nu}^{\mu}+\sigma_{\nu}^{\mu}\right.\nonumber \\
 &  & \qquad\qquad\quad\quad+r^{2}f\left(br\right)\partial_{r}F\left(br\right)b\sigma_{\nu}^{\mu}\nonumber \\
 &  & \left.\qquad\qquad\quad\quad-\frac{1}{2}\frac{d}{f\left(br\right)\left(br\right)^{d}}u^{\mu}u^{\rho}\partial_{\rho}u_{\nu}\right)\label{eq:first order extrinsic curvature}\end{eqnarray}
}{\large \par}

{\large \begin{equation}
K^{\left(0+1\right)}=K^{\left(0\right)}\label{eq:first order trace of extrinsic curvature}\end{equation}
}{\large \par}

{\large We define the quantities with the superscript: $^{\left(0\right)},\,^{\left(1\right)},\,^{\left(2\right)}$
to be quantities that have no derivative, only one derivative or two
derivative terms respectively. Now we insert (\ref{eq:first order extrinsic curvature})
and (\ref{eq:first order trace of extrinsic curvature}) to Gauss-Codazzi
equations (\ref{eq:Gauss-Codazzi equations with vanishing matter T_{ab}})
and get the constraint equations to the second order:}{\large \par}

{\large \begin{eqnarray}
K_{\nu|\mu}^{\mu\left(0+1\right)}-K_{,\nu}^{\left(0+1\right)} & = & K_{\nu|\mu}^{\mu\left(0\right)}-K_{,\nu}^{\left(0\right)}+K_{\nu|\mu}^{\mu\left(1\right)}-K_{,\nu}^{\left(1\right)}\nonumber \\
 & = & \partial_{\mu}K_{\nu}^{\mu\left(0\right)}+\bar{\Gamma}_{\mu\sigma}^{\mu\left(1\right)}K_{\nu}^{\sigma\left(0\right)}-\bar{\Gamma}_{\mu\nu}^{\sigma\left(1\right)}K_{\sigma}^{\mu\left(0\right)}-\partial_{\nu}K^{\left(0\right)}+\nonumber \\
 &  & +\partial_{\mu}K_{\nu}^{\mu\left(1\right)}+\bar{\Gamma}_{\mu\sigma}^{\mu\left(1\right)}K_{\nu}^{\sigma\left(1\right)}+\bar{\Gamma}_{\mu\sigma}^{\mu\left(2\right)}K_{\nu}^{\sigma\left(0\right)}\nonumber \\
 &  & -\bar{\Gamma}_{\mu\nu}^{\sigma\left(1\right)}K_{\sigma}^{\mu\left(1\right)}-\bar{\Gamma}_{\mu\nu}^{\sigma\left(2\right)}K_{\sigma}^{\mu\left(0\right)}\nonumber \\
 & = & \partial_{\mu}K_{\nu}^{\mu\left(0\right)}+\bar{\Gamma}_{\mu\sigma}^{\mu\left(1\right)}K_{\nu}^{\sigma\left(0\right)}-\bar{\Gamma}_{\mu\nu}^{\sigma\left(1\right)}K_{\sigma}^{\mu\left(0\right)}-\partial_{\nu}K^{\left(0\right)}+\nonumber \\
 &  & +\frac{d}{2\left(bR\right)^{d}\sqrt{\left|f\left(bR\right)\right|}}\frac{2b}{d}\left(\partial_{\mu}\sigma_{\nu}^{\mu}-\left(d-1\right)A_{\mu}\sigma_{\nu}^{\mu}\right)=0\nonumber \\
\label{eq:Gauss-Codazzi R=00003Dconst 2 order}\end{eqnarray}
where the bar over the Christoffel symbols represents the Christoffel
symbols which have been calculated with respect to the metric $h_{\mu\nu}$
of the hyper-surface $\Phi$. The fifth line is the LHS of equation
(\ref{eq:Gauss-Codazzi R=00003Dconst 1 order}).}{\large \par}

{\large Projecting (\ref{eq:Gauss-Codazzi R=00003Dconst 2 order})
along $u^{\nu}$ we get,}{\large \par}

{\large \begin{eqnarray}
 &  & \frac{d}{2\left(bR\right)^{d}\sqrt{\left|f\left(bR\right)\right|}}\left(\partial_{\mu}u^{\mu}-\left(d-1\right)u^{\nu}\partial_{\nu}\ln b-\frac{2b}{d}\sigma_{\alpha\beta}\sigma^{\alpha\beta}\right)=0\Rightarrow\nonumber \\
 &  & \Rightarrow\partial_{\mu}u^{\mu}-\left(d-1\right)u^{\nu}\partial_{\nu}\ln b=\frac{2b}{d}\sigma_{\alpha\beta}\sigma^{\alpha\beta}\label{eq:constraint equation 1st order 1}\end{eqnarray}
where we used the relation,}{\large \par}

{\large \begin{eqnarray*}
-\partial_{\mu}\sigma_{\nu}^{\mu}u^{\nu} & = & \sigma_{\nu}^{\mu}\partial_{\mu}u^{\nu}\\
 & = & \sigma_{\nu}^{\mu}P^{\mu\alpha}P_{\beta}^{\nu}\partial_{\alpha}u^{\beta}\\
 & = & \sigma_{\nu}^{\mu}P^{\mu\alpha}P_{\beta}^{\nu}\frac{1}{2}\left(\partial_{\alpha}u^{\beta}+\partial_{\beta}u^{\alpha}\right)\\
 & = & \sigma_{\nu}^{\mu}\left(P_{\mu}^{\alpha}P_{\beta}^{\nu}\frac{1}{2}\left(\partial_{\alpha}u^{\beta}+\partial_{\beta}u^{\alpha}\right)-P_{\mu}^{\nu}\frac{\partial_{\alpha}u^{\alpha}}{d-1}\right)\\
 & = & \sigma_{\alpha\beta}\sigma^{\alpha\beta}\end{eqnarray*}
}{\large \par}

{\large Projecting (\ref{eq:Gauss-Codazzi R=00003Dconst 2 order})
along $P_{\sigma}^{\nu}$,}{\large \par}

{\large \begin{eqnarray}
 &  & \frac{d}{2\left(bR\right)^{d}\sqrt{\left|f\left(bR\right)\right|}}\nonumber \\
 &  & \times\left(-u^{\mu}\partial_{\mu}u_{\sigma}+P_{\sigma}^{\nu}\partial_{\nu}\ln b+\frac{2b}{d}P_{\sigma}^{\nu}\left(\partial_{\mu}\sigma_{\nu}^{\mu}-\left(d-1\right)A_{\mu}\sigma_{\nu}^{\mu}\right)\right)=0\Rightarrow\nonumber \\
 &  & \Rightarrow u^{\mu}\partial_{\mu}u_{\sigma}-P_{\sigma}^{\nu}\partial_{\nu}\ln b=\frac{2b}{d}P_{\sigma}^{\nu}\left(\partial_{\mu}\sigma_{\nu}^{\mu}-\left(d-1\right)A_{\mu}\sigma_{\nu}^{\mu}\right)\label{eq:constraint equations 1st order 2}\end{eqnarray}
}{\large \par}

{\large We can see that our results are in agreement with the equations
of motion (\ref{eq:the equations of motion}) of a conformal fluid
to the second order in derivative expansion, that is, up to first
order terms of the stress tensor (\ref{eq:stress tensor for conformal fluid 2nd order}).}{\large \par}

\section{{\large Non-Constant Radial Location}}

{\large We can generalize our discussion to non-constant hyper-surfaces,
i.e.: $R=R\left(x\right)$. In order to do so, we need to use the
property in which we can explicitly calculate the $r$ derivatives
of tensors which have been calculated in the bulk. For instance, the
constraint equations, $E_{\mu}^{r}=R_{\mu}^{r}=0$ can have the following
form:}{\large \par}

{\large \begin{equation}
R_{\mu}^{r}=\sum_{i}g_{i}\left(r,b\right)h_{i\mu}\left(\partial u\left(x\right),\partial b\left(x\right),\partial\partial u\left(x\right),\partial\partial b\left(x\right),higher\, terms\right)\label{eq:generalize expression for constraints}\end{equation}
}{\large \par}

{\large Note that in this expression there are no derivatives of $r$
because we can calculate them explicitly. However, derivatives of
$x^{\mu}$ can not be calculated explicitly, because they are operating
on arbitrary functions $u^{\mu}\left(x\right),\, b\left(x\right)$.}{\large \par}

{\large For the second order constraint equations, we have no higher
terms in the brackets of (\ref{eq:generalize expression for constraints})
and the answer for constant $R$ was calculated above by Gauss-Codazzi
equations (\ref{eq:Gauss-Codazzi R=00003Dconst 2 order}). From (\ref{eq:guess-codazzi contracted after getting  rid of the metric})
and the claim in chapter 4 (\ref{eq:Claim}) we have:}{\large \par}

{\large \begin{equation}
\left.\frac{1}{\mathcal{N}}R_{\nu}^{r\left(1+2\right)}\right|_{r=R}=K_{\nu|\mu}^{\mu\left(0+1\right)}-K_{,\nu}^{\left(0+1\right)}=g_{1}\left(R\right)h_{1\nu}=0\end{equation}
where:}{\large \par}

{\large \begin{equation}
g_{1}\left(R\right)=\frac{d}{2\left(bR\right)^{d}\sqrt{\left|f\left(bR\right)\right|}}\end{equation}
}{\large \par}

{\large \begin{eqnarray}
h_{1\nu} & = & \left(\vphantom{\frac{2b}{d}}2du_{\nu}u^{\mu}\partial_{\mu}\ln b-u^{\mu}\partial_{\mu}u_{\nu}\right.\nonumber \\
 &  & -u_{\nu}\partial_{\mu}u^{\mu}+\left(d+1\right)\partial_{\nu}\ln b-dP_{\nu}^{\mu}\partial_{\mu}\ln b\nonumber \\
 &  & \left.+\frac{2b}{d}\left(\partial_{\mu}\sigma_{\nu}^{\mu}-\left(d-1\right)A_{\mu}\sigma_{\nu}^{\mu}\right)\right)\end{eqnarray}
and}{\large \par}

{\large \begin{eqnarray}
\mathcal{N} & = & \sqrt{\left|g^{ab}\partial_{a}\left(r-r\left(x^{\alpha}\right)\right)\partial_{b}\left(r-r\left(x^{\alpha}\right)\right)\right|}\end{eqnarray}
}{\large \par}

{\large Therefore, the sum in (\ref{eq:generalize expression for constraints})
has only one term: $g_{i}\left(R\right)=0\;\forall i\geq2$. However,
the latter is supposed to be true for $\forall R=const$. Thus: $g_{i}\left(r\right)=0\;\forall i\geq2$.
Therefore, we obtain the constraint equations for the second order
in a derivative expansion}%
\footnote{Because $g_{1}\left(R\right)\neq0$, we can divide the equations:
$g_{1}\left(R\right)h_{1\nu}=0$ by $g_{1}\left(R\right)$.%
}{\large{} which are the same for any non-null hyper-surface}%
\footnote{However, the constraint equations for $f\left(bR\right)=0$, which
is in the case of a null hyper-surface, hold true and are still the
same as in the non-null case \cite{key-2}.

The case in which $R=0$ is not define.%
}{\large .}{\large \par}

{\large \newpage{}}{\large \par}

\chapter{{\LARGE Event vs Apparent Horizon}}

{\large In this chapter we review the different notions of black hole
horizons. We will focus our discussion on two distinct horizons, namely,
the apparent and the event horizons. The last section in this chapter
is devoted to calculating the constraint equations from the second
order bulk metric projected on the different horizons.}{\large \par}

\section{{\large Event Horizon}}

{\large One approach to define the unique hyper-surface that defines
the boundary of a black hole is the causal approach. In this approach
one defines the unique hyper-surface, which is called the }\emph{\large event
horizon,}{\large{} as the boundary of the casual past of future null
infinity \cite{key-16}. This mathematical definition can be put in
a more physical language by the following statement: the event horizon
is the boundary of the region in space-time from which nothing can
ever escape.}{\large \par}

{\large We should note here that the future null infinity is a time-like
hyper-surface. That is because we deal with an asymptotically AdS
space-time, for which the future directed null geodesics (which is
the future null infinity in the definition) end on a time-like hyper-surface.}{\large \par}

{\large One should also note that by definition the event horizon
is a null hyper-surface which is generated by null geodesics.}{\large \par}

{\large We would like to describe a method for finding the location
of the event horizon. This method requires the knowledge of the late
time generators of the event horizon, which means the location of
the event horizon in late time. The method goes as follow: evolve
the geodesic equation backwards in time and use the knowledge of the
late time generators as the future boundary condition.}{\large \par}

{\large For our metric, which describes a dynamical black hole which
is dual to viscous fluid flow, we can assume that after a while the
fluid relaxes, which means that we have a stationary black hole that
is described by the metric (\ref{eq:static metrics}). The location
of the event horizon for the metric is well known and is $r_{EH}^{stat}=\frac{1}{b}$.
Since we are working in a small Knudsen number i.e., derivative expansion,
we can find the location of the event horizon order by order \cite{key-4}:}{\large \par}

{\large \begin{equation}
r_{EH}\left(x^{\alpha}\right)=\sum_{l=0}^{\infty}r_{EH}^{\left(l\right)}\left(x^{\alpha}\right)\end{equation}
}{\large \par}

{\large We require that the normal vector $\ell^{a}$ to the event
horizon hyper-surface $\mathcal{S}_{EH}$ , defined by:}{\large \par}

{\large \begin{equation}
\mathcal{S}_{EH}\left(r,x\right)=r-r_{EH}\left(x\right),\;\ell^{a}=g^{ab}\partial_{b}\,\mathcal{S}_{EH}\end{equation}
will be a null vector $\ell^{a}\ell_{a}=0$ and we get the equation,}{\large \par}

{\large \begin{equation}
g^{rr}-2\partial_{\mu}r_{EH}g^{r\mu}+\partial_{\mu}r_{EH}\partial_{\nu}r_{EH}g^{\mu\nu}=0\end{equation}
which can be solved algebraically by using the initial condition for
the location of the stationary event horizon: $r_{EH}^{\left(0\right)}=\frac{1}{b}$
. The solution up to the second order,}{\large \par}

{\large \begin{equation}
r_{EH}\left(x^{\alpha}\right)=\frac{1}{b\left(x\right)}+b\left(x\right)\left(h_{1}\sigma_{\alpha\beta}\sigma^{\alpha\beta}+h_{2}\omega_{\alpha\beta}\omega^{\alpha\beta}+h_{3}\mathcal{R}\right)\label{eq:Event Horizon Location}\end{equation}
with}{\large \par}

{\large \begin{eqnarray}
h_{1} & = & \frac{2\left(d^{2}+d-4\right)}{d^{2}\left(d-1\right)\left(d-2\right)}-\frac{K_{2}\left(1\right)}{d\left(d-1\right)}\nonumber \\
h_{2} & = & -\frac{d+2}{2d\left(d-2\right)}\label{eq:event horizon parameters}\\
h_{3} & = & -\frac{1}{d\left(d-1\right)\left(d-2\right)}\nonumber \end{eqnarray}
}{\large \par}

\begin{figure}
\begin{centering}
{\large \includegraphics[scale=0.3]{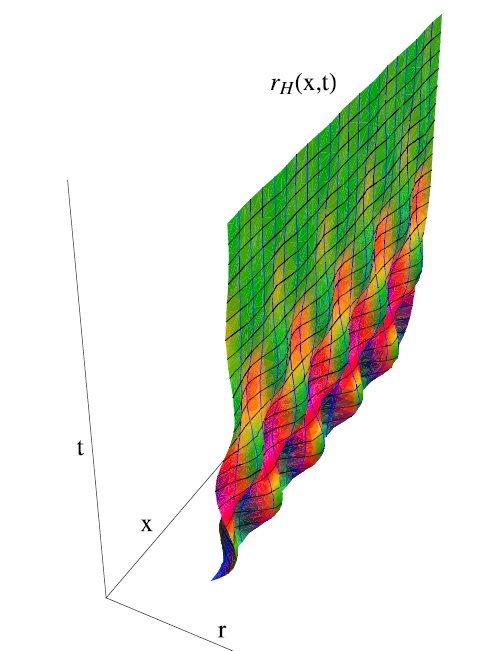}}
\par\end{centering}{\large \par}

{\large \caption{{\large \label{figure 3}}The event horizon $r=r_{EH}(x^{\mu})$ is
sketched as a function of the time $t$ and one of the spatial coordinates
$x$ (the other $d-2$ spatial coordinates are suppressed). (Taken
from \cite{key-9}).}
}
\end{figure}

{\large Note that the solution is dependent on the boundary coordinates
alone, which means that for a constant $x^{\mu}$, i.e., a tube along
the radial direction, we have a fixed solution and that in late time
we get our starting condition that is the static solution (figure
\ref{figure 3}).}{\large \par}

{\large }%
{\large \par}

\section{{\large Apparent Horizon}}

{\large We would now like to describe a black hole by relating to
its strong gravitational fields instead of its causal structure. The
latter definition led us to the introduction of the event horizon
hyper-surface in the previous section. In this section we show the
major drawback in that definition, which leads us to a new hyper-surface,
the apparent horizon \cite{key-17}.}{\large \par}

\begin{figure}
\begin{centering}
{\large \includegraphics[scale=0.25]{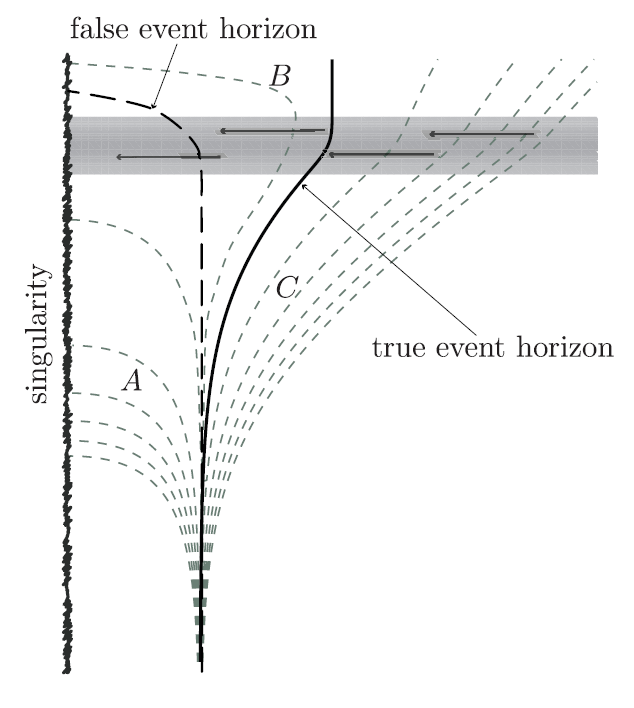}}
\par\end{centering}{\large \par}

{\large \caption{{\large \label{figure 4}}A schematic demonstrating the non-local
nature of event horizon evolution for a spherically symmetric space-time
with the angular dimensions suppressed. Horizontal location measures
the radius of the associated spherical shell while time is (roughly)
vertical. The shaded gray region represents infalling null dust. (Taken
from \cite{key-5}).}
}
\end{figure}

{\large In order to find the location of the event horizon, one needs
to know the entire future evolution of space-time, which means that
one needs to wait infinite time in order to find the location of the
event horizon. To illustrate the problem, we will present the example
given in \cite{key-5}. Imagine a Schwarzschild black hole which has
been irradiated by an infalling shell of null dust (figure \ref{figure 4}).
This is represented by the gray rectangle. The inward and outward
null directions are tangent to the light gray dashed lines. Note that
the inward null direction is horizontal. In order to find the event
horizon, one needs to track the evolution of radial null geodesics.
If they fall into the singularity, they are inside the black hole
and if they are heading outward, even after the null dust passes,
then they are outside the black hole. If one wants to provide the
location of the event horizon without having any prior knowledge of
the arrival of the null dust, one would state that it is located at
the hard dashed line, which is clearly a false statement. One can
realize this by examining the radial null geodesic B, which seems
to be escaping from the black hole, but ends at the singularity. By
a close examination one sees that the location of the event horizon
is the hard line. One can see how it curves a bit in anticipation
for the null dust to arrive, as if it \textquotedblleft{}knows\textquotedblright{}
that it is coming.}{\large \par}

{\large }%
{\large \par}

{\large We will now explain how a new perspective on a dynamical black
hole avoids the causal problem. This perspective is, as mentioned
before, the strong gravitational field approach. In order to describe
this approach we will give an example found in \cite{key-17,key-5}.
Imagine a spherical shell, in a four dimensional world, which emits
light and then stops. The light front null geodesics will be divided
into two families that move in two null directions: the outgoing null
direction and the ingoing null direction. We will denote the tangent
to the light ray null vectors, which are perpendicular to the spherical
shell surface, by $\ell^{a}$ and $n^{a}$ respectively. The outgoing
light front will increase in cross-sectional area and diverge, while
the ingoing light front will decrease in the cross-sectional area
and converge. In a more mathematical way, if we define the }\emph{\large expansion
parameter $\theta$}{\large{} \cite{key-13} as the fractional rate
of change (per unit affine-parameter distance) of the congruence's
(a family of geodesics that does not intersect each other) cross-sectional
area:}{\large \par}

{\large \begin{equation}
\theta=\frac{1}{\delta A}\frac{d}{d\lambda}\delta A\end{equation}
where $\lambda$ is an affine-parameter, and $\delta A$ is the infinitesimal
cross-sectional area measured in the purely transverse directions}%
\footnote{The geodesics are crossing the cross-sectional area $\delta A$ orthogonally. %
}{\large , then we can describe the convergence and divergence of the
light front by the following:}{\large \par}

{\large }%
\begin{figure}
\begin{centering}
{\large \includegraphics[scale=0.3]{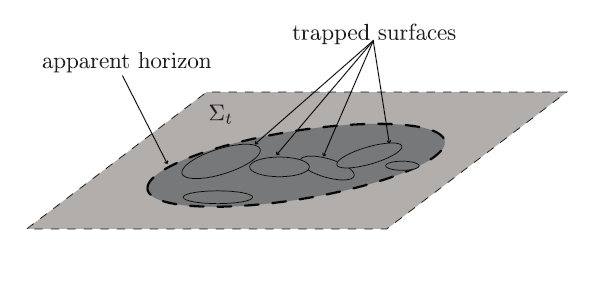}}
\par\end{centering}{\large \par}

{\large \caption{{\large \label{figure 5}}An \textquotedblleft{}instant\textquotedblright{}
$\Sigma_{t}$ along with some of its trapped surfaces (small black
circles), the associated trapped region (dark gray) and the apparent
horizon (thick dashed line). (Taken from \cite{key-5}).}
}
\end{figure}
{\large \par}

{\large \begin{equation}
\theta_{\left(\ell\right)}>0,\quad\theta_{\left(n\right)}<0\end{equation}
where the sub-script implies to which light front the expansion is
referring to. A positive expansion means that the cross-sectional
area will increase while negative expansion means that it will decrease.}{\large \par}

{\large Now we are ready to introduce the notion of a trapped surface:
A }\emph{\large trapped surface}{\large{} in $d+1$ space-time dimensions
is a space-like hyper-surface of $d-1$ dimensions, which, by definition,
has two null normals: the outgoing null normal $\ell^{a}$ and the
ingoing null normal $n^{a}$, where the expansion in both directions
is negative.}{\large \par}

{\large \begin{equation}
\theta_{\left(\ell\right)}<0,\quad\theta_{\left(n\right)}<0\end{equation}
}{\large \par}

{\large The hyper-surface that surrounds all of the trapped surfaces
is called the }\emph{\large apparent horizon}{\large }%
\footnote{The apparent horizon is a special case of the so called \emph{marginally
trapped surface} reviewed in \cite{key-18} %
}{\large{} (figure \ref{figure 5}) and defined mathematically by:}{\large \par}

{\large \begin{equation}
\theta_{\left(\ell\right)}=0,\quad\theta_{\left(n\right)}<0\end{equation}
}{\large \par}

\begin{figure}
\begin{centering}
{\large \includegraphics[scale=0.3]{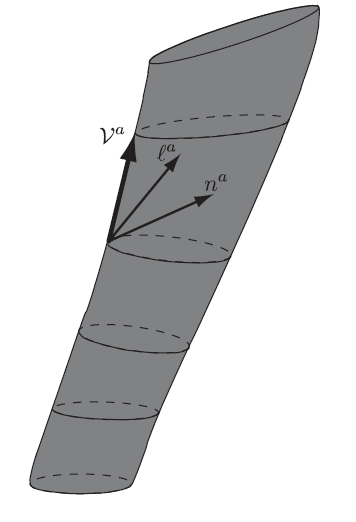}}
\par\end{centering}{\large \par}

{\large \caption{{\large \label{figure 6}}A schematic of an $d$-tube $\Delta$ with
compact foliation hyper-surfaces $S_{\lambda}$ along with the outward
and inward pointing null normals to those hyper-surfaces. $\nu^{a}$
is the future-pointing tangent to $\Delta$ that is simultaneously
normal to the $S_{\lambda}$. (Taken from \cite{key-5}).}
}
\end{figure}

{\large Note to the condition $\theta_{\left(\ell\right)}=0$. It
does not mean that if matter comes into the black-hole the apparent
horizon location will not change (increase)}%
\footnote{The expansion of the apparent horizon is related the expansion in
the direction of the evolution vector $\nu^{a}$ (\ref{eq:evolution vector}),
which is:

\[
\theta_{\left(\nu\right)}=\theta_{\left(\ell\right)}-C\theta_{\left(n\right)}=-C\theta_{\left(n\right)}>0\]
}{\large . If we let the apparent horizon evolve in time, we get a
co-dimension one tube (which means the dimensions of the tube is $d$).
The tube is built in the following manner: given a foliation of space-time
with $d-1$ hyper-surfaces $S_{\lambda}$ which have two null normal
vectors: $\ell^{a}$ and $n^{a}$ as defined above, we connect them
with an evolution vector field $\nu^{a}$ that maps every point in
$S_{\lambda}$ to $S_{\lambda+1}$ to create the $d$-dimensional
tube $\Delta$ (figure \ref{figure 6}). The evolution vector is perpendicular
to each of the hyper-surface $S_{\lambda}$ and tangent to the $d$-dimensional
tube $\Delta$. If the $S_{\lambda}$ are the apparent horizon hyper-surfaces,
then we say that$\Delta$ is a }\emph{\large time evolved apparent
horizon}{\large }%
\footnote{From now on we will refer to the time evolved apparent horizon as
the apparent horizon.%
}{\large .}{\large \par}

{\large }%
{\large \par}

{\large We will now explain about the geometric quantities that we
defined. First we state that we cross normalized the null normal vectors,}{\large \par}

{\large \begin{equation}
\ell^{a}n_{a}=-1\label{eq:cross normalization}\end{equation}
}{\large \par}

{\large Moreover, we present their normalization:}{\large \par}

{\large \begin{equation}
\ell^{a}\rightarrow f\ell^{a}\quad n^{a}\rightarrow\frac{1}{f}n^{a}\label{eq:normalization f}\end{equation}
where $f$ is an arbitrary function.}{\large \par}

{\large Then we define the evolution vector to be normal to the $S_{\lambda}$
hyper-surfaces, so it will be built from the two null vectors}%
\footnote{Because we are using null vectors, we can always choose a normalization
that will give us this form of evolution vector. %
}{\large{} $\ell^{a}$ and $n^{a}$:}{\large \par}

{\large \begin{equation}
\nu^{a}=\ell^{a}-Cn^{a}\label{eq:evolution vector}\end{equation}
for some function C. Note that if $C>0$, then we get space-like hyper-surface,
if $C=0$, we get a null hyper-surface and if $C<0$, we get a time-like
hyper-surface. }{\large \par}

{\large Now we can present $m^{a}$, the normal to the $d$-tube hyper-surface
$\Delta$ vector, which is perpendicular to the evolution vector and
we get:}{\large \par}

{\large \begin{equation}
m^{a}=\ell^{a}+Cn^{a}\end{equation}
}{\large \par}

{\large Because $m^{a}$ is normal to the hyper-surface $\Delta$,
which can be presented by the hyper-surface equation $\phi\left(r,x\right)=0$,
we will choose to define it as:}{\large \par}

{\large \begin{equation}
m^{a}=g^{ab}\partial_{b}\phi\end{equation}
which is similar to what we did in (\ref{eq:normal vector}) but without
having to normalize it}%
\footnote{Note that $m^{a}$ can be space-like, time-like and even null.%
}{\large .}{\large \par}

{\large We pause here and return to the event horizon description.
The event horizon can also be described by the $d$-tube description
and of course its evolution vector will be with $C=0$, so we actually
see that both the normal vector to the hyper-surface $\Delta$ (the
event horizon) and the evolution vector are the same:}{\large \par}

{\large \begin{equation}
m^{a}=\nu^{a}=\ell^{a}\end{equation}
where $\ell^{a}$ is defined in (\ref{eq:null vector defintion})}%
\footnote{In the rest of the section we will set the normalization of $f$ in
(\ref{eq:null vector defintion}) to one.%
}{\large{} }{\large \par}

{\large We now specify some more geometric quantities and we introduce
the metric for the hyper-surface $S_{\lambda}$, as well: }{\large \par}

{\large \begin{equation}
q_{ab}=g_{ab}+\ell_{a}n_{b}+\ell_{b}n_{a}\label{eq:projector onto the d-1 hyper-surface}\end{equation}
}{\large \par}

{\large Note that this metric has two zero eigenvalues.}{\large \par}

{\large We can associate the expansion parameters to the geometric
quantities using the following definition:}{\large \par}

{\large \begin{equation}
\theta_{\left(\ell\right)}=q^{ab}\nabla_{a}\ell_{b},\quad\theta_{\left(n\right)}=q^{ab}\nabla_{a}n_{b}\label{eq:expansion defintion with the q metric}\end{equation}
}{\large \par}

{\large Note that $q^{ab}$ is defined as in (\ref{eq:projector onto the d-1 hyper-surface}),
but with upper indices and it is not the inverse of $q_{ab}$}%
\footnote{One can think of $q^{ab}$ as a projection onto the $S_{\lambda}$
hyper-surfaces, which in (\ref{eq:expansion defintion with the q metric})
restricts $\nabla_{a}\ell_{b}$ and $\nabla_{a}n_{b}$ to the hyper-surfaces.
The expansion $\theta$ is the trace after the restriction.%
}{\large .}{\large \par}

\begin{figure}
\begin{centering}
{\large \includegraphics[scale=0.3]{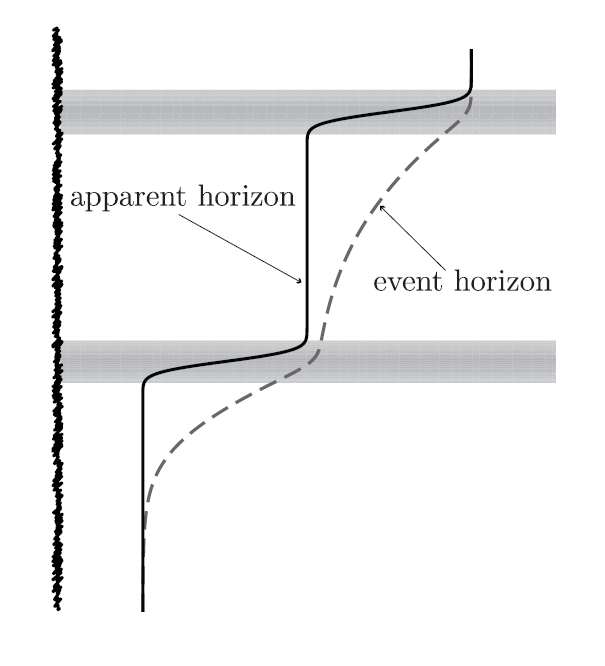}}
\par\end{centering}{\large \par}

{\large \caption{{\large \label{figure 7}}A simulation similar to that of Figure 4
though this time two distinct shells fall into the black hole. Both
the apparent and event horizons are plotted. (Taken from \cite{key-5}).}
}{\large \par}

\end{figure}

{\large Now we are ready to return to the causal problem (the global
definition of the event horizon) presented earlier and see whether
we can provide a solution to it. One can see (figure \ref{figure 7})
that the apparent horizon changes only when the null dust comes in,
as opposed to the event horizon that \textquotedblleft{}anticipates\textquotedblright{}
the arrival of the null dust, and curves even before it arrives. Therefore,
we have found a solution to the problem. Note that the apparent horizon
is located inside the black hole, which means that it lies within
the event horizon. Moreover, the apparent and the event horizons are
the same in the stationary case; one can see this by looking at their
location after the two null dusts pass (figure \ref{figure 7}).}{\large \par}

{\large }%
{\large \par}

{\large We want to find the location of the apparent horizon in the
same manner that we did for the event horizon, by using a derivative
expansion:}{\large \par}

{\large \begin{equation}
r_{AH}\left(x^{\alpha}\right)=\sum_{l=0}^{\infty}r_{AH}^{\left(l\right)}\left(x^{\alpha}\right)\label{eq:expansion to be detrmined of AH}\end{equation}
}{\large \par}

{\large From the requirement $\theta_{\left(\ell\right)}=0$ we can
find the location order by order.}{\large \par}

{\large However, in order to do so we need to foliate space-time with
$d-1$ hyper-surfaces $S_{\lambda}$. We define those surfaces by
specifying the two null vectors to the surfaces. The symmetry of space-time
singles out the $S_{\lambda}$ hyper-surfaces of constant $r$ and
constant {}``time'' propagation. The latter is an arbitrary combination
of $x^{\mu}$. The outward null normal $\ell^{a}$ and the inward
null normal $n^{a}$ are chosen to be:}{\large \par}

{\large \begin{equation}
\ell^{a}=\mathcal{B}\left(\frac{\partial}{\partial r}\right)^{a}+\mathcal{N^{\mu}}\left(\frac{\partial}{\partial x^{\mu}}\right)^{a}\end{equation}
}{\large \par}

{\large \begin{equation}
n^{a}=-\left(\frac{\partial}{\partial r}\right)^{a}\end{equation}
}{\large \par}

{\large We can determined them by the relation (\ref{eq:cross normalization})
and the fact that they are both null vectors and we get}{\large \par}

{\large \begin{equation}
\mathcal{N^{\mu}}=u^{\mu}+\mathcal{T}^{\mu},\quad\mathcal{T^{\mu}}u_{\mu}=0\end{equation}
}{\large \par}

{\large \begin{eqnarray}
\mathcal{B} & = & -\mathcal{N^{\mu}}\mathcal{V}_{\mu}\end{eqnarray}
to the second order, where $\mathcal{T^{\mu}}$ is composed from one
or two derivatives of the $d$-velocity or the temperature, so we
have freedom in determining this quantity. We therefore require that
this description be suited to describe the event horizon, because
we know that the apparent horizon coincides with the event horizon
in the stationary case. In order to do so, we require that the null
outward vector in the $x^{\mu}$ direction resembles the null outward
vector in the $x^{\mu}$ direction that defines the event horizon.
This requirement determines $\mathcal{T^{\mu}}$, and we get:}{\large \par}

{\large \begin{eqnarray}
\mathcal{T^{\mu}} & = & \left(1-\frac{1}{br}\right)bu^{\lambda}\partial_{\lambda}u^{\mu}-\frac{1}{r^{2}}\frac{1}{d-2}P^{\mu\rho}\left(\partial_{\lambda}-\left(d-3\right)A_{\lambda}\right)\omega_{\rho}^{\lambda}+\nonumber \\
 &  & +\left(\frac{1}{r^{2}}\left(2L\left(br\right)+\frac{1}{d-2}\right)-\frac{2b^{2}}{d}\right)P^{\mu\rho}\left(\partial_{\lambda}-\left(d-1\right)A_{\lambda}\right)\sigma_{\rho}^{\lambda}+\nonumber \\
 &  & +\left(1-\frac{1}{br}\right)\frac{2b}{r}F\left(br\right)\sigma^{\mu\nu}u^{\lambda}\partial_{\lambda}u_{\nu}\end{eqnarray}
}{\large \par}

{\large Now we are ready to find the expansion in the outward null
direction:}{\large \par}

{\large \begin{eqnarray}
\theta_{\left(\ell\right)} & = & q^{ab}\nabla_{a}\ell_{b}\nonumber \\
 & = & \nabla_{a}\ell^{a}+\ell^{a}n^{b}\nabla_{a}\ell_{b}\nonumber \\
 & = & \partial_{a}\ell^{a}+\Gamma_{ab}^{a}\ell^{b}+\ell^{a}n^{b}\left(\partial_{a}\ell_{b}-\Gamma_{ab}^{c}\ell_{c}\right)\nonumber \\
 & = & \partial_{r}\ell^{r}+\partial_{\mu}\ell^{\mu}+2\Gamma_{r\mu}^{r}\ell^{\mu}+\Gamma_{\nu r}^{\nu}\ell^{r}+\Gamma_{\nu\mu}^{\nu}\ell^{\mu}+\ell^{\nu}\Gamma_{\nu r}^{\mu}\ell_{\mu}\nonumber \\
 & = & -\frac{\left(d-1\right)}{2r\left(br\right)^{d}}\left[r^{2}\left(1-\left(br\right)^{d}\right)-\left(\frac{2\left(br\right)^{d}}{\left(d-2\right)}+\frac{1}{2}\right)\omega_{\alpha\beta}\omega^{\alpha\beta}+\right.\nonumber \\
 &  & +\left.\left(\frac{2\left(br\right)^{d-2}}{d-2}-\frac{K_{2}\left(br\right)}{d-1}\right)\left(br\right)^{2}\sigma_{\alpha\beta}\sigma^{\alpha\beta}\right]+\frac{\mathcal{R}}{2r\left(d-2\right)}\nonumber \\
 &  & -\frac{r^{2}\left(1-\left(br\right)^{d}\right)}{2\left(br\right)^{d}}\left(-b^{2}\partial_{r}K_{1}\left(br\right)\sigma_{\alpha\beta}\sigma^{\alpha\beta}+\frac{1}{r^{3}}\omega_{\alpha\beta}\omega^{\alpha\beta}\right)\qquad\label{eq:expansion to second order}\end{eqnarray}
}{\large \par}

{\large Inserting (\ref{eq:expansion to be detrmined of AH}) to (\ref{eq:expansion to second order})
and using $\theta_{\left(\ell\right)}=0$, we find order by order
the location of the apparent horizon:}{\large \par}

{\large \begin{equation}
r_{AH}\left(x^{\alpha}\right)=\frac{1}{b\left(x\right)}+b\left(x\right)\left(\widetilde{h}_{1}\sigma_{\alpha\beta}\sigma^{\alpha\beta}+h_{2}\omega_{\alpha\beta}\omega^{\alpha\beta}+h_{3}\mathcal{R}\right)\end{equation}
}{\large \par}

{\large \begin{equation}
\widetilde{h}_{1}=h_{1}-\frac{4}{d^{2}\left(d-1\right)}\end{equation}
where $h_{1},\: h_{2},\: h_{3}$ are defined in (\ref{eq:event horizon parameters}).}{\large \par}

{\large Note that up to the first order in derivative expansion, the
event and apparent horizons coincide.}{\large \par}

\section{{\large Event and Apparent Horizons Constraint Equations}}

{\large We are now well equipped to check if the constraint equations
for the event and apparent horizons are the same up to the third order
in a derivative expansion. The work that was done in \cite{key-2}
showed that the event horizon, to the first order in derivative expansion
of the metric, exhibits the same constraint equations to the second
order in derivative expansion}%
\footnote{The constraint equations always have one derivative more than the
derivative expansion of the metric%
}{\large , as in the case of non-null hyper-surface (\ref{eq:constraint equation 1st order 1}),
(\ref{eq:constraint equations 1st order 2}). As we mentioned above,
the event and apparent horizons differ only in the second order in
derivative expansion. Therefore, the constraint equations up to the
second order are the same in all of the hyper-surfaces mentioned above.
Recall our statement about the ability to get the same constraint
equations from different hyper-surfaces}%
\footnote{Actually for hyper-surfaces of the form $\phi=r-r\left(x^{\alpha}\right)$.%
}{\large , we see that it turns out to be right up to first order in
the metric.}{\large \par}

{\large In order to check this statement for the event and apparent
horizons to the second order in derivative expansion of the metric,
let us look directly at the constraint equations i.e.: $E_{\mu}^{r}$
and calculate their difference while they are evaluated on the event
and apparent horizons hyper-surfaces:}{\large \par}

{\large \begin{eqnarray}
EH-AH & = & \left.E_{\mu}^{r}\right|_{r=r_{EH}}-\left.E_{\mu}^{r}\right|_{r=r_{AH}}=\left.R_{\mu}^{r}\right|_{r=r_{EH}}-\left.R_{\mu}^{r}\right|_{r=r_{AH}}\label{eq:Difference between EH and AH the easy way}\end{eqnarray}
}{\large \par}

{\large Bear in mind that the difference in $R_{\mu}^{r}$ between
the event and apparent horizons to the third order will only come
from the evaluation of $R_{\mu}^{r}$ in the bulk to the first order
and from the location of the horizons to the second order. Recall
that the difference between the horizons location is only in the second
order.}{\large \par}

{\large Now we present the calculation of the tensor $R_{\mu}^{r}$
to the first order in the bulk:}{\large \par}

{\large \begin{equation}
R_{\mu}^{r\left(1\right)}=\frac{1}{2}r\frac{d}{\left(br\right)^{d}}\left(\partial_{\mu}\ln b+du^{\rho}\partial_{\rho}\ln b-u_{\mu}\partial_{\alpha}u^{\alpha}-u^{\nu}\partial_{\nu}u_{\mu}\right)\label{eq:Ricci tensor 1st order}\end{equation}
}{\large \par}

{\large From Einstein's equations (\ref{eq:Einstein's equations})
we know that $R_{\mu}^{r\left(1\right)}=0$ for any $r$. Therefore,
the expression in the brackets in (\ref{eq:Ricci tensor 1st order})
is zero. This expression is just the constraint equations we found
in (\ref{eq:constraint equation zeroth order 1}) and (\ref{eq:constraint equation zeroth order 2}).}{\large \par}

{\large Now we are ready to calculate (\ref{eq:Difference between EH and AH the easy way})
to the third order:}{\large \par}

{\large \begin{eqnarray}
EH-AH & = & \left.R_{\mu}^{r\left(1\right)}\right|_{r=r_{EH}}-\left.R_{\mu}^{r\left(1\right)}\right|_{r=r_{AH}}\nonumber \\
 & = & \left(\frac{1}{2}r_{EH}\frac{d}{\left(br_{EH}\right)^{d}}-\frac{1}{2}r_{AH}\frac{d}{\left(br_{AH}\right)^{d}}\right)\left(\partial_{\mu}\ln b\right.\nonumber \\
 &  & \left.+du^{\rho}\partial_{\rho}\ln b-u_{\mu}\partial_{\alpha}u^{\alpha}-u^{\nu}\partial_{\nu}u_{\mu}\right)\nonumber \\
 & = & 0\end{eqnarray}
where we get the fourth line by using the constraint equation to the
first order. }{\large \par}

{\large There is another method to get this result. This method is
more cumbersome than what we presented here, however, it poses challenging
computational techniques; therefore, we presented it in the appendix.}{\large \par}

{\large \newpage{}}{\large \par}

\chapter{{\LARGE Hydrodynamic Quantities}}

\section{Entropy Current}

{\large In \cite{key-4} the authors introduce a boundary entropy
current obtained from the event horizon by using the area of the black
hole event horizon. By using the connection between the area of a
hyper-surface to the square root of the metric determinant of that
hyper-surface, and by using a pull-back along ingoing null geodesics
to that area, we get the entropy current on the boundary}%
\footnote{Of course not every hyper-surface is adequate. We require that the
hyper-surface admits the area increase theorem, that gives us a natural
candidate: the event horizon. However, it is not the only hyper-surface
with that property.%
}{\large . In order to do so, we split the boundary coordinate $x^{\mu}$
to $\left(\nu,x^{i}\right)$ and we continue to use the same coordinates
on the hyper-surface $\Phi=r-r\left(x\right)$. We will get, as in
the $d$-tube description, a space-like hyper-surfaces $\Sigma_{\nu}$
which is a constant $\nu$ slice. This, as well, will divide the components
of $m^{\mu}$ into $\left(m^{\nu},m^{i}\right)$}%
\footnote{Here $\nu$ does not represent $d$ different coordinates, but a specific
coordinate. These notations were also used in \cite{key-4}. %
}{\large . After we split space-time, we define the metric on the space-like
hyper-surface $\Sigma_{\nu}$ by the restriction of the line element
to be only on that hyper-surface:}{\large \par}

{\large \begin{equation}
ds^{2}=g_{ab}dx^{a}dx^{b}=\widetilde{q}_{ij}dx^{j}dx^{j}\end{equation}
}{\large \par}

{\large Then by using the procedure done in \cite{key-4}we get:}{\large \par}

{\large \begin{equation}
J^{\mu}=\frac{\sqrt{\widetilde{q}}}{4}\frac{m^{\mu}}{m^{\nu}}\label{eq:entropy current definition}\end{equation}
where $\widetilde{q}$ is the determinant of $\widetilde{q}_{ij}$.
It can be shown that this form of entropy current is indeed covariant
\cite{key-4}. Also the authors in \cite{key-4} found the entropy
current for the event horizon hyper-surface, and \cite{key-1} reproduced
it to arbitrary $d$-dimensions:}{\large \par}

{\large \begin{eqnarray}
4b^{d-1}J_{EH}^{\mu} & = & u^{\mu}+b^{2}u^{\mu}\left[A_{1}\sigma_{\alpha\beta}\sigma^{\alpha\beta}+A_{2}\omega_{\alpha\beta}\omega^{\alpha\beta}+A_{3}\mathcal{R}\right]+\nonumber \\
 & + & b^{2}\left[B_{1}D_{\lambda}\sigma^{\mu\lambda}+B_{2}D_{\lambda}\omega^{\mu\lambda}\right]\end{eqnarray}
with,}{\large \par}

{\large \begin{eqnarray}
A_{1} & = & \frac{2}{d^{2}}\left(d+2\right)-\frac{K_{1}\left(1\right)d+K_{2}\left(1\right)}{d},\quad A_{2}=-\frac{1}{2d},\quad B_{2}=\frac{1}{d-2},\nonumber \\
B_{1} & = & -2A_{3}=\frac{2}{d\left(d-2\right)}\label{eq:event horizon entropy coefficients}\end{eqnarray}
}{\large \par}

{\large In order to evaluate the the entropy current to the second
order in derivative expansion for the apparent horizon, we calculate
the difference between the two entropy currents:}{\large \par}

{\large \begin{equation}
J_{EH}^{\mu}-J_{AH}^{\mu}=\left.\frac{\sqrt{\widetilde{q}}}{4}\frac{m^{\mu}}{m^{\nu}}\right|_{r=r_{EH}}-\quad\left.\frac{\sqrt{\widetilde{q}}}{4}\frac{m^{\mu}}{m^{\nu}}\right|_{r=r_{AH}}\end{equation}
}{\large \par}

{\large Because $m^{\mu}$ for the event horizon and apparent horizon
is the same (\ref{eq:Identities for the difference calculation}),
we need to evaluate the difference in $\sqrt{\widetilde{q}}$ on the
horizons. The difference to the order we are seeking will come from
inputting the location of the horizon to the second order into the
zero order of the metric determinant,}{\large \par}

{\large \begin{equation}
\sqrt{\widetilde{q}}^{\left(0\right)}=r^{d-1}\end{equation}
and we get the difference between the entropy currents to be:}{\large \par}

{\large \begin{eqnarray}
J_{EH}^{\mu}-J_{AH}^{\mu} & = & \frac{1}{4}\frac{m^{\mu}}{m^{\nu}}\left(\left.\sqrt{\widetilde{q}}\right|_{r=r_{EH}}-\left.\sqrt{\widetilde{q}}\right|_{r=r_{AH}}\right)\nonumber \\
 & = & \frac{1}{4}\frac{m^{\mu}}{m^{\nu}}\left(\left.r^{d-1}\right|_{r=r_{EH}}-\left.r^{d-1}\right|_{r=r_{AH}}\right)\nonumber \\
 & = & \frac{\left(d-1\right)}{4}\frac{m^{\mu}}{m^{\nu}}b^{-\left(d-2\right)}\left(\left.r^{\left(2\right)}\right|_{r=r_{EH}}-\left.r^{\left(2\right)}\right|_{r=r_{AH}}\right)\nonumber \\
 & = & \frac{m^{\mu}}{m^{\nu}}\frac{b^{-\left(d-3\right)}}{d^{2}}\sigma_{\alpha\beta}\sigma^{\alpha\beta}\end{eqnarray}
}{\large \par}

{\large Therefore, we see that the entropy current for the apparent
horizon is,}{\large \par}

{\large \begin{eqnarray}
4b^{d-1}J_{AH}^{\mu} & = & u^{\mu}+b^{2}u^{\mu}\left[\widetilde{A}_{1}\sigma_{\alpha\beta}\sigma^{\alpha\beta}+A_{2}\omega_{\alpha\beta}\omega^{\alpha\beta}+A_{3}\mathcal{R}\right]+\nonumber \\
 & + & b^{2}\left[B_{1}D_{\lambda}\sigma^{\mu\lambda}+B_{2}D_{\lambda}\omega^{\mu\lambda}\right]\end{eqnarray}
with,}{\large \par}

{\large \begin{eqnarray}
\widetilde{A}_{1} & = & A_{1}-\frac{4}{d^{2}}\end{eqnarray}
and the other coefficients are the same as the coefficients of the
event horizon entropy current (\ref{eq:event horizon entropy coefficients}).}{\large \par}

{\large We now have to check if this entropy current be suitable to
describe the notion of entropy, which means we need to answer the
question: does the entropy current never decreases? }{\large \par}

{\large In order to answer this question \cite{key-4} calculated
the divergence of the entropy current and required the non-negativeness
of the expression:}{\large \par}

{\large \begin{equation}
D_{\mu}J^{\mu}\geq0\label{eq:non-negativeness of the DJ}\end{equation}
}{\large \par}

{\large They did it with arbitrary coefficients $A_{1},\: A_{2},\: A_{3},\: B_{1},\: B_{2}$
and found that in order that (\ref{eq:non-negativeness of the DJ})
applies, the following connection should be valid:}{\large \par}

{\large \begin{equation}
B_{1}=2A_{3}\end{equation}
}{\large \par}

{\large This connection, of course, is holds true for the event horizon
and also for the apparent horizon. Therefore, we can say that it is
possible to ascribe a different entropy current to the boundary by
the apparent horizon hyper-surface.}{\large \par}

\section{Temperature }

{\large Besides the entropy current, we can relate another thermodynamic
quantity of the fluid to gravity, that is the temperature.}{\large \par}

{\large The temperature of the black hole can be related to the surface
gravity of the event horizon by $T_{EH}=\frac{\kappa}{2\pi}$. We
use a generalized definition for the surface gravity shown in \cite{key-5}:}{\large \par}

{\large \begin{equation}
\kappa_{\nu}=-n_{b}\nu^{a}\nabla_{a}\ell^{b}=-n_{b}\ell^{a}\nabla_{a}\ell^{b}+Cn_{b}n^{a}\nabla_{a}\ell^{b}=\kappa_{\ell}+C\kappa_{n}\end{equation}
to calculate the surface gravity for different hyper-surfaces. }%
{}{\large \par}

{\large The result for a hyper-surface located at $r=\frac{1}{b}+r^{\left(2\right)}$
(}$r^{\left(2\right)}${\large{} has second order derivatives) is:}{\large \par}

{\large \begin{eqnarray}
\kappa_{\nu} & = & \kappa_{\ell}\nonumber \\
 & = & \frac{d}{2b}-\frac{\partial_{\lambda}u^{\lambda}}{d-1}\nonumber \\
 & - & \frac{d}{2}\left(d-3\right)r^{\left(2\right)}+\partial_{r}\left[\frac{1}{2\left(br\right)^{d}}\left(\frac{1}{2}\omega_{\alpha\beta}\omega^{\alpha\beta}\right.\right.\nonumber \\
 &  & \left.\left.+\left(br\right)^{2}K_{2}\left(br\right)\frac{\sigma_{\alpha\beta}\sigma^{\alpha\beta}}{d-1}\right)\right]\end{eqnarray}
and by inserting the location of the event horizon or the apparent
horizon, we can get their associated surface gravity respectively.
These quantities may relate to different notion of temperature fields
by dividing them by $2\pi$.}{\large \par}

{\large In order to set the surface gravity to be equal, up to $2\pi$,
to the temperature field of the conformal fluid we set the normalization
of (\ref{eq:normalization f}) in such a way that we get: $T_{fluid}=\frac{\kappa_{\ell}}{2\pi}$.
The surface gravity transforms in the following way:}{\large \par}

{\large \begin{equation}
\kappa_{\ell}^{\prime}=f\left(\kappa_{\ell}+\frac{d}{d\lambda}\ln f\right)\end{equation}
where $f$ is the normalization in (\ref{eq:normalization f}), $\kappa_{\ell}^{\prime}$
is the new surface gravity and $\lambda$ is an arbitrary parameter}%
\footnote{Note that this is the transformation of the change of the parameter
$\lambda$, because if $\lambda\rightarrow\lambda^{*}$, then $\ell^{*a}=\frac{d\lambda}{d\lambda^{*}}\ell^{a}\equiv f\,\ell^{a}$.%
}{\large .}{\large \par}

{\large Under this change of the normalization $f$, we need to check
how the other quantities that we defined transform and if this transformation
changes our results regarding the location of the apparent horizon
and the entropy current.}{\large \par}

{\large The location of the apparent horizon depends on how the expansion
parameter transforms:}{\large \par}

{\large \begin{equation}
\theta_{\left(\ell\right)}^{\prime}=f\,\theta_{\left(\ell\right)}\end{equation}
and we can see that it does not change the location of the apparent
horizon. We can see this by trying to find the {}``new'' location
of the apparent horizon by setting $\theta_{\left(\ell\right)}^{\prime}=0$.
It means that if $f\neq0$, which is the trivial case, then $\theta_{\left(\ell\right)}=0$
which is exactly the {}``old'' location of the apparent horizon.}{\large \par}

{\large By the definition of the entropy current (\ref{eq:entropy current definition})
one can see that it is invariant to the normalization transformation.
\newpage{}}{\large \par}

\chapter{{\LARGE Conclusions}}

{\large In this thesis we presented a method and different techniques
to evaluate properties of an out-of-equilibrium black hole that has
a dual description as a conformal out-of-equilibrium relativistic
fluid, which \textquotedblleft{}lives\textquotedblright{} on the boundary
of AdS space-time. We set the stage and presented the derivative expansion
for both the relativistic fluid and for the black-hole. We presented
a way to compute the Einstein constraint equations for different hyper-surfaces.
The constraint equations turned out to be just the conservation equations
of the fluid. We found, as expected, that from any hyper-surface which
is located at $r=r\left(x\right)$, we get the same conservation equations.
This implies that maybe one can describe a hydrodynamic duality, not
just from the quantities that are related to a certain hyper-surface,
namely the event horizon, but also to quantities related to other
hyper-surfaces. We suggested the apparent horizon as a possible candidate,
and calculated from the apparent horizon area the entropy current
and from its surface gravity the temperature for a conformal fluid.
Our work poses a few future questions. The first is: which hyper-surface
describes correctly the conformal fluid properties? The second is:
how can we use these results for the description of the QGP (Quark
Gluon Plasma)? }{\large \par}

{\large Our solution for the difference between the location and the
entropy current of the event and apparent horizons is in agreement
with the boost invariant Bjorken flow which was done in \cite{key-5,key-19}.
Only recently we came across a similar work \cite{key-20} that finds
the entropy for the apparent horizon. They used a slightly different
approach than what we did to define the entropy current for the apparent
horizon. However, it is in complete agreement with our results.}{\large \par}

{\large Moreover, a recent work \cite{key-19} suggests that the apparent
horizon is a better candidate to describe the dual fluid flow at the
boundary. This work examines the exact solution of the conformal soliton
flow, which results in the divergence of the area of the event horizon,
while the apparent horizon stays finite and constant. This suggests
that the corresponding entropy of the fluid should be described by
the apparent horizon. }{\large \par}

\chapter*{Acknowledgements}

{\large I would like to thank my thesis supervisor, Prof. Yaron Oz,
who introduced me to this fascinating world. His guidance and support
from the initial to the final levels enabled me to develop an understanding
of the subject. }{\large \par}

{\large I would also like to express my deep appreciation to Yasha
Neiman for valuable discussions.}{\large \par}

\appendix

\chapter{{\LARGE Notation}}

{\large We refer to lower Roman indices that start from the beginning
of the alphabet $a,b\lyxmathsym{\ldots},h$ as the whole $d+1$ space-time
coordinates (bulk coordinate). The Greek indices $\mu,\nu...$ refer
to the $d$ dimensions hyper-surface of the fluid that \textquotedbl{}lives\textquotedbl{}
on the boundary of the bulk, which are just the indices of the CFT.
The lower Roman indices that start from $i,j...$ refer to the $d-1$
spatial directions of the fluid.}{\large \par}

{\large We commonly use different metrics, which we will specify here:}{\large \par}
\begin{itemize}
\item {\large $g_{ab}$ - is the $d+1$-dimensional bulk metric. We use
this metric to lower and raise tensors with bulk indices that are
defined in the whole space-time.}{\large \par}
\item {\large $h_{ab}$ - is a $d+1$-dimensional metric that is restricted
to a certain hyper-surface. We use this metric to lower and raise
tensors with bulk indices that are defined on a certain hyper-surface.}{\large \par}
\item {\large $h_{\mu\nu}$ - is a $d$-dimensional metric of a hyper-surface.
We get this metric by a projection of $g_{ab}$ into th}\textbf{\large e
}{\large hyper-surface. We use this metric to lower and raise tensors
with Greek indices that are defined on a certain hyper-surface.}{\large \par}
\item {\large $g_{\mu\nu}$ - is a $d$-dimensional boundary metric. In
our calculation $g_{\mu\nu}$ is the flat metric $\eta_{\mu\nu}$.
Only in the section that deals with the Weyl formalism, do we let
it be also a curved metric.}{\large \par}
\item {\large $\eta_{\mu\nu}$ - is the $d$-dimensional Minkowski boundary
metric, with the signature convention of \textquotedbl{}more plus
signs\textquotedbl{} $\left(-,+,+,...\right)$. This is the metric
of the fluid in the boundary and we will lower and raise tensors with
boundary Greek indices with that metric, i.e., $u^{\mu}=\eta^{\mu\nu}u_{\nu}$.}{\large \par}
\item {\large $q_{ab}$ - is a $d+1$-dimensional metric, which defines
the $d-1$ space-like hyper-surfaces that create the $d$-dimensional
tube. This metric projects the bulk tensor onto the $d-1$ hyper-surface.
We use this metric to lower and raise tensors that we want to restrict
into the $d-1$ hyper-surface and that are written in the bulk indices,
$a,b,...$.}{\large \par}
\item {\large $\widetilde{q}_{ij}$ - is a $d-1$-dimensional metric, which
defines the $d-1$ space-like hyper-surfaces that create the $d$-dimensional
tube. We use this metric to lower and raise tensors that are restricted
to the $d-1$ hyper-surface and have $i,j...$ indices.}{\large \par}
\end{itemize}
{\large We remind the reader of the basic tensors used commonly in
this thesis:}{\large \par}

{\large The }\emph{\large Riemann curvature tensor}{\large{} defined
below for bulk vector field $A^{a}$,}{\large \par}

{\large \begin{equation}
\nabla_{c}\nabla_{d}A^{a}-\nabla_{d}\nabla_{c}A^{a}=R_{bcd\quad}^{a}A^{b}\end{equation}
}{\large \par}

{\large \begin{equation}
R_{bcd}^{a}\equiv\Gamma_{bd,c}^{a}-\Gamma_{cb,d}^{a}+\Gamma_{ce}^{a}\Gamma_{db}^{e}-\Gamma_{de}^{a}\Gamma_{cb}^{e}\end{equation}
}{\large \par}

{\large Additionally we define the }\emph{\large Ricci tensor}{\large{}
and the }\emph{\large Ricci scalar}{\large{} by,}{\large \par}

{\large \begin{equation}
R_{bd}=R_{bad}^{a}\end{equation}
}{\large \par}

{\large \begin{equation}
R=g^{ab}R_{ab}\end{equation}
}{\large \par}

\chapter{{\LARGE Detailed Calculations }}

{\large For the hyper-surface $\Phi=r-R$, where $R$ is constant,
we have the basis vectors written in (\ref{eq:basis for constant R hyper-surface}).
Then the metric on the surface, to any order, is:}{\large \par}

{\large \begin{equation}
h_{\mu\nu}=g_{ab}e_{\mu}^{a}e_{\nu}^{b}=g_{\mu\nu}\end{equation}
}{\large \par}

{\large The inverse metrics to the zero and the first order are:}{\large \par}

{\large \begin{equation}
h_{\left(0\right)}^{\mu\nu}=\frac{1}{R^{2}}P^{\mu\nu}-\frac{1}{R^{2}f\left(bR\right)}u^{\mu}u^{\nu}\end{equation}
}{\large \par}

{\large \begin{eqnarray}
h_{\left(0+1\right)}^{\mu\nu} & = & h_{\left(0\right)}^{\mu\nu}-\frac{2bF\left(bR\right)}{R^{2}}\sigma^{\mu\nu}-\frac{2}{d-1}\frac{1}{R^{3}f\left(bR\right)^{2}}\partial_{\lambda}u^{\lambda}u^{\mu}u^{\nu}\nonumber \\
 & + & \frac{1}{R^{3}f\left(bR\right)}u^{\lambda}\partial_{\lambda}\left(u^{\mu}u^{\nu}\right)\end{eqnarray}
}{\large \par}

{\large The Christoffel symbol, with respect to the metric of the
hyper-surface $\Phi$ to the first order, is:}{\large \par}

{\large \begin{eqnarray}
\bar{\Gamma}_{\mu\nu}^{\sigma\left(1\right)} & =\frac{1}{2}h^{\sigma\lambda} & \left(h_{\mu\lambda,\nu}+h_{\nu\lambda,\mu}-h_{\mu\nu,\lambda}\right)\nonumber \\
 & = & \frac{1}{2\left(bR\right)^{d}}\left[\frac{1}{2}u_{(\mu}\partial_{\nu)}u^{\sigma}+du_{\mu}u_{\nu}P^{\sigma\lambda}\partial_{\lambda}\ln b\right.\nonumber \\
 &  & \quad\quad\quad\quad-P^{\sigma\lambda}\partial_{\lambda}\left(u_{\mu}u_{\nu}\right)+\frac{1}{f\left(bR\right)}u^{\sigma}\left(\frac{1}{2}\partial_{(\nu}u_{\mu)}\right.\nonumber \\
 &  & \quad\quad\quad\quad-\frac{1}{2}du_{(\mu}\partial_{\nu)}\ln b-du^{\lambda}u_{\mu}u_{\nu}\partial_{\lambda}\ln b\nonumber \\
 &  & \left.\left.\quad\quad\quad\quad+u^{\lambda}\partial_{\lambda}\left(u_{\mu}u_{\nu}\right)\vphantom{\frac{1}{2}}\right)\vphantom{\frac{1}{2}}\right]\end{eqnarray}
}{\large \par}

{\large \begin{equation}
\bar{\Gamma}_{\mu\nu}^{\mu\left(2\right)}=\frac{1}{d-1}\frac{1}{Rf\left(bR\right)}\left(\partial_{\rho}u^{\rho}\partial_{\nu}\ln f\left(bR\right)-\partial_{\nu}\partial_{\rho}u^{\rho}\right)\end{equation}
}{\large \par}

{\large \begin{eqnarray}
\bar{\Gamma}_{\mu\nu}^{\sigma\left(2\right)}u^{\mu}u_{\sigma} & = & \frac{1}{2}\frac{1}{Rf\left(bR\right)}\left(u_{\nu}u^{\rho}\partial_{\rho}u^{\lambda}\partial_{\lambda}f\left(bR\right)+2\frac{\partial_{\nu}\partial_{\rho}u^{\rho}}{d-1}\right.\nonumber \\
 &  & \quad\quad\quad\quad\quad-2u^{\rho}\partial_{\rho}u^{\lambda}\partial_{\nu}u_{\lambda}-2\frac{\partial_{\rho}u^{\rho}}{d-1}\partial_{\nu}\ln f\left(bR\right)\nonumber \\
 &  & \quad\quad\quad\quad\quad+\left(1-f\left(bR\right)\right)u^{\rho}\left(\partial_{\rho}u^{\lambda}\partial_{\nu}u_{\lambda}\right.\nonumber \\
 &  & \left.\left.\quad\quad\quad\quad\quad-u_{\nu}\partial_{\rho}u^{\lambda}u^{\mu}\partial_{\mu}u_{\lambda}-\partial_{\rho}u^{\lambda}\partial_{\lambda}u_{\nu}\right)\vphantom{\frac{\partial_{\rho}u^{\rho}}{d-1}}\right)\end{eqnarray}
}{\large \par}

{\large We will specify the detailed calculation of the Christoffel
symbols for the entire bulk space-time.}{\large \par}

{\large The Christoffel symbols that we can calculate in full to all
orders, of the full bulk metric $g_{ab}$, are:\begin{eqnarray}
\Gamma_{rr}^{\nu} & = & \frac{1}{2}g^{\nu a}\left(g_{ra,r}+g_{ra,r}-g_{rr,a}\right)\nonumber \\
 & = & \frac{1}{2}g^{\nu r}\left(g_{rr,r}+g_{rr,r}-g_{rr,r}\right)\nonumber \\
 &  & +\frac{1}{2}g^{\nu\lambda}\left(g_{r\lambda,r}+g_{r\lambda,r}-g_{rr,\lambda}\right)=0\end{eqnarray}
\begin{eqnarray}
\Gamma_{rr}^{r} & = & \frac{1}{2}g^{ra}\left(g_{ra,r}+g_{ra,r}-g_{rr,a}\right)\nonumber \\
 & = & \frac{1}{2}g^{rr}\left(g_{rr,r}+g_{rr,r}-g_{rr,r}\right)\nonumber \\
 &  & +\frac{1}{2}g^{r\lambda}\left(g_{r\lambda,r}+g_{r\lambda,r}-g_{rr,\lambda}\right)=0\end{eqnarray}
}{\large \par}

{\large The Christoffel symbols that are needed to calculate the Ricci
tensor, with respect to the metric $g_{ab}^{\left(0+1\right)}=g_{ab}^{\left(0\right)}+g_{ab}^{\left(1\right)}$,
to the first order are:}{\large \par}

{\large \begin{eqnarray}
\Gamma_{\mu\sigma}^{\nu\left(0+1\right)} & = & \frac{1}{2}g^{\nu a}\left(g_{\mu a,\sigma}+g_{\sigma a,\mu}-g_{\mu\sigma,a}\right)\nonumber \\
 & = & \frac{1}{2}g^{\nu r}\left(g_{\mu r,\sigma}+g_{\sigma r,\mu}-g_{\mu\sigma,r}\right)\nonumber \\
 &  & +\frac{1}{2}g^{\nu\lambda}\left(g_{\mu\lambda,\sigma}+g_{\sigma\lambda,\mu}-g_{\mu\sigma,\lambda}\right)\nonumber \\
 & = & \frac{1}{2}u^{\nu}\left(-2\partial_{(\mu}u_{\sigma)}-\partial_{r}\left(-r^{2}f\left(br\right)u_{\mu}u_{\sigma}+r^{2}P_{\mu\sigma}\right)\right)\nonumber \\
 &  & +\frac{1}{2}\left[2\left(1-f\left(br\right)\right)u_{(\mu}\partial_{\sigma)}u^{\nu}\right.\nonumber \\
 &  & \left.-P^{\nu\rho}\partial_{\rho}\left(-f\left(br\right)u_{\mu}u_{\sigma}+P_{\mu\sigma}\right)\right]\nonumber \\
 &  & -b\partial_{r}\left(r^{2}F\left(br\right)\right)u^{\nu}\sigma_{\mu\sigma}-\frac{\partial_{\lambda}u^{\lambda}}{d-1}u^{\nu}u_{\mu}u_{\sigma}\nonumber \\
 &  & +\frac{1}{2}u^{\nu}u^{\rho}\partial_{\rho}\left(u_{\mu}u_{\sigma}\right)-\frac{1}{2r}\partial_{r}\left(r^{2}f\left(br\right)\right)u_{\mu}u_{\sigma}u^{\rho}\partial_{\rho}u^{\nu}\nonumber \\
 &  & +u^{\rho}\partial_{\rho}u^{\nu}P_{\mu\sigma}\end{eqnarray}
}{\large \par}

{\large \begin{eqnarray}
\Gamma_{\mu r}^{\nu\left(0+1\right)} & = & \frac{1}{2}g^{\nu a}\left(g_{\mu a,r}+g_{ra,\mu}-g_{\mu r,a}\right)\nonumber \\
 & = & \frac{1}{2}g^{\nu r}\left(g_{\mu r,r}+g_{rr,\mu}-g_{\mu r,r}\right)+\frac{1}{2}g^{\nu\lambda}\left(g_{\mu\lambda,r}+g_{r\lambda,\mu}-g_{\mu r,\lambda}\right)\nonumber \\
 & = & \frac{1}{r}P_{\mu}^{\nu}+\frac{1}{2r^{2}}+\omega_{\mu}^{\:\nu}+b\partial_{r}\left(F\left(br\right)\right)\sigma_{\mu}^{\nu}\end{eqnarray}
}{\large \par}

{\large \begin{eqnarray}
\Gamma_{\mu\nu}^{r\left(0+1\right)} & = & \frac{1}{2}g^{ra}\left(g_{\mu a,\nu}+g_{\nu a,\mu}-g_{\mu\nu,a}\right)\nonumber \\
 & = & \frac{1}{2}g^{rr}\left(g_{\mu r,\nu}+g_{\nu r,\mu}-g_{\mu\nu,r}\right)+\frac{1}{2}g^{r\lambda}\left(g_{\mu\lambda,\nu}+g_{\nu\lambda,\mu}-g_{\mu\nu,\lambda}\right)\nonumber \\
 & = & \frac{1}{2}r^{2}f\left(br\right)\left(-2\partial_{(\mu}u_{\nu)}-\partial_{r}\left(-r^{2}f\left(br\right)u_{\mu}u_{\nu}+r^{2}P_{\mu\nu}\right)\right)\nonumber \\
 &  & +\frac{1}{2}r^{2}\left[2\partial_{(\nu}\left(f\left(br\right)u_{\mu)}\right)\right.\nonumber \\
 &  & \left.-2\partial_{(\nu}u_{\mu)}-u^{\rho}\partial_{\rho}\left(-f\left(br\right)u_{\mu}u_{\nu}+P_{\mu\nu}\right)\right]\nonumber \\
 &  & +r^{2}f\left(br\right)\left(\vphantom{\frac{\partial_{\lambda}u^{\lambda}}{d-1}}-b\partial_{r}\left(r^{2}F\left(br\right)\right)\sigma_{\mu\nu}\right.\nonumber \\
 &  & \left.-\frac{\partial_{\lambda}u^{\lambda}}{d-1}u_{\mu}u_{\nu}+\frac{1}{2}u^{\lambda}u_{\lambda}\left(u_{\mu}u_{\nu}\right)\right)\nonumber \\
 &  & -\frac{\partial_{\lambda}u^{\lambda}}{d-1}r\partial_{r}\left(r^{2}f\left(br\right)\right)u_{\mu}u_{\nu}+2r^{2}\frac{\partial_{\lambda}u^{\lambda}}{d-1}P_{\mu\nu}\end{eqnarray}
}{\large \par}

{\large \begin{eqnarray}
\Gamma_{\mu r}^{r\left(0+1\right)} & = & \frac{1}{2}g^{ra}\left(g_{\mu a,r}+g_{ra,\mu}-g_{\mu r,a}\right)\nonumber \\
 & = & \frac{1}{2}g^{rr}\left(g_{\mu r,r}+g_{rr,\mu}-g_{\mu r,r}\right)\nonumber \\
 &  & +\frac{1}{2}g^{r\lambda}\Gamma_{\mu\nu}^{r\left(0+1\right)}\left(g_{\mu\lambda,r}+g_{r\lambda,\mu}-g_{\mu r,\lambda}\right)\nonumber \\
 & = & \frac{1}{2}\partial_{r}\left(r^{2}f\left(br\right)\right)u_{\mu}-\frac{\partial_{\lambda}u^{\lambda}}{d-1}u_{\mu}\end{eqnarray}
The Christoffel symbols that are needed to calculate the expansion
parameter $\theta_{\left(\ell\right)}$, with respect to the metric
$g_{ab}^{\left(0+1+2\right)}=g_{ab}^{\left(0\right)}+g_{ab}^{\left(1\right)}+g_{ab}^{\left(2\right)}$
to the second order, are:\begin{eqnarray}
\Gamma_{\mu\nu}^{r\left(0+1\right)} & = & \frac{1}{2}g^{ra}\left(g_{\mu a,\nu}+g_{\nu a,\mu}-g_{\mu\nu,a}\right)\nonumber \\
 & = & \frac{1}{2}g^{rr}\left(g_{\mu r,\nu}+g_{\nu r,\mu}-g_{\mu\nu,r}\right)+\frac{1}{2}g^{r\lambda}\left(g_{\mu\lambda,\nu}+g_{\nu\lambda,\mu}-g_{\mu\nu,\lambda}\right)\nonumber \\
 & = & \frac{1}{2}r^{2}f\left(br\right)\left(-2\partial_{(\mu}u_{\nu)}-\partial_{r}\left(-r^{2}f\left(br\right)u_{\mu}u_{\nu}+r^{2}P_{\mu\nu}\right)\right)\nonumber \\
 &  & +\frac{1}{2}r^{2}\left[2\partial_{(\nu}\left(f\left(br\right)u_{\mu)}\right)-2\partial_{(\nu}u_{\mu)}\right.\nonumber \\
 &  & \left.-u^{\rho}\partial_{\rho}\left(-f\left(br\right)u_{\mu}u_{\nu}+P_{\mu\nu}\right)\right]\nonumber \\
 &  & +r^{2}f\left(br\right)\left(\vphantom{\frac{\partial_{\lambda}u^{\lambda}}{d-1}}-b\partial_{r}\left(r^{2}F\left(br\right)\right)\sigma_{\mu\nu}\right.\nonumber \\
 &  & \left.-\frac{\partial_{\lambda}u^{\lambda}}{d-1}u_{\mu}u_{\nu}+\frac{1}{2}u^{\lambda}u_{\lambda}\left(u_{\mu}u_{\nu}\right)\right)\nonumber \\
 &  & -\frac{\partial_{\lambda}u^{\lambda}}{d-1}r\partial_{r}\left(r^{2}f\left(br\right)\right)u_{\mu}u_{\nu}+2r^{2}\frac{\partial_{\lambda}u^{\lambda}}{d-1}P_{\mu\nu}\end{eqnarray}
\begin{eqnarray}
\Gamma_{r\mu}^{r\left(0+1+2\right)} & = & \frac{1}{2}g^{ra}\left(g_{\mu a,r}+g_{ra,\mu}-g_{\mu r,a}\right)\nonumber \\
 & = & \frac{1}{2}g^{rr}\left(g_{\mu r,r}+g_{rr,\mu}-g_{\mu r,r}\right)+\frac{1}{2}g^{r\lambda}\left(g_{\mu\lambda,r}+g_{r\lambda,\mu}-g_{\mu r,\lambda}\right)\nonumber \\
 & = & \frac{1}{2}\left(u^{\nu}-\mathfrak{\left(G^{-1}\right)^{\nu\sigma}}\mathcal{V_{\sigma}}\right)\left(-2\partial_{[\mu}u_{\nu]}+\partial_{r}\left(-2u_{(\mu}\mathcal{V}_{\nu)}+\mathfrak{G_{\mu\nu}}\right)\right)\nonumber \\
 & = & -u_{\mu}\frac{\partial_{\lambda}u^{\lambda}}{d-1}-u_{\mu}\partial_{r}\left[\frac{1}{2\left(br\right)^{d}}\left(\vphantom{\frac{\sigma_{\alpha\beta}\sigma^{\alpha\beta}}{d-1}}r^{2}\left(1-\left(br\right)^{d}\right)\right.\right.\nonumber \\
 &  & \left.\left.-\frac{1}{2}\omega_{\alpha\beta}\omega^{\alpha\beta}-\left(br\right)^{2}K_{2}\left(br\right)\frac{\sigma_{\alpha\beta}\sigma^{\alpha\beta}}{d-1}\right)\vphantom{\frac{1}{2\left(br\right)^{d}}}\right]\nonumber \\
 &  & +\frac{1}{2r}u^{\lambda}\partial_{\lambda}u_{\sigma}P_{\mu}^{\lambda}\partial_{\lambda}u^{\sigma}-br\partial_{r}F\left(br\right)u^{\lambda}\partial_{\lambda}u_{\sigma}\sigma_{\mu}^{\sigma}\nonumber \\
 &  & -\frac{1}{r}\frac{2}{d-2}P_{\mu}^{\rho}\left(\partial_{\lambda}-\left(d-3\right)A_{\lambda}\right)\omega_{\rho}^{\lambda}\nonumber \\
 &  & +\left(\frac{2L\left(br\right)}{r\left(br\right)^{d-2}}-\frac{1}{2}\partial_{r}\left(\frac{2L\left(br\right)}{\left(br\right)^{d-2}}\right)+\frac{1}{r}\frac{1}{d-2}\right)\nonumber \\
 &  & \times P_{\mu}^{\rho}\left(\partial_{\lambda}-\left(d-1\right)A_{\lambda}\right)\sigma_{\rho}^{\lambda}\nonumber \\
 &  & -\frac{1}{2r}u^{\lambda}\partial_{\lambda}u^{\sigma}\partial_{\sigma}u_{\mu}\end{eqnarray}
}{\large \par}

{\large \begin{eqnarray}
\Gamma_{\nu r}^{\nu\left(0+1+2\right)} & = & \frac{1}{2}g^{\nu a}\left(g_{\nu a,r}+g_{ra,\nu}-g_{\nu r,a}\right)\nonumber \\
 & = & \frac{1}{2}g^{\nu r}\left(g_{\nu r,r}+g_{rr,\nu}-g_{\nu r,r}\right)+\frac{1}{2}g^{\nu\lambda}\left(g_{\nu\lambda,r}+g_{r\lambda,\nu}-g_{\nu r,\lambda}\right)\nonumber \\
 & = & \mathfrak{\frac{1}{2}\left(G^{-1}\right)^{\nu\sigma}}\partial_{r}\mathfrak{G_{\nu\sigma}}\nonumber \\
 & = & \frac{1}{r}\left(d-1\right)-b^{2}\partial_{r}K_{1}\left(br\right)\sigma_{\alpha\beta}\sigma^{\alpha\beta}\nonumber \\
 &  & +\frac{b^{2}}{r}P^{\alpha\beta}u^{\rho}\partial_{\rho}\sigma_{\alpha\beta}\partial_{r}\left(H_{2}\left(br\right)-H_{1}\left(br\right)\right)\nonumber \\
 &  & +\frac{1}{r^{3}}\omega_{\alpha\beta}\omega^{\alpha\beta}\end{eqnarray}
}{\large \par}

{\large \begin{eqnarray}
\Gamma_{\nu\mu}^{\nu\left(0+1+2\right)} & = & \frac{1}{2}g^{\nu a}\left(g_{\nu a,\mu}+g_{\mu a,\nu}-g_{\nu\mu,a}\right)\nonumber \\
 & = & \frac{1}{2}g^{\nu r}\left(g_{\nu r,\mu}+g_{\mu r,\nu}-g_{\nu\mu,r}\right)+\frac{1}{2}g^{\nu\lambda}\left(g_{\nu\lambda,\mu}+g_{\mu\lambda,\nu}-g_{\nu\mu,\lambda}\right)\nonumber \\
 & = & \frac{1}{2}\left(u^{\nu}-\left(\mathcal{\mathfrak{G}}^{-1}\right)^{\nu\sigma}\mathcal{V}_{\sigma}\right)\left(-2\partial_{(\mu}u_{\nu)}-\partial_{r}\left(-2u_{(\mu}\mathcal{V}_{\nu)}+\mathfrak{G_{\mu\nu}}\right)\right)\nonumber \\
 &  & +\frac{1}{2}\left(\mathcal{\mathfrak{G}}^{-1}\right)^{\nu\sigma}\partial_{\mu}\left(-2u_{(\sigma}\mathcal{V}_{\nu)}+\mathfrak{G_{\sigma\nu}}\right)\nonumber \\
 & = & u_{\mu}\frac{\partial_{\lambda}u^{\lambda}}{d-1}+u_{\mu}\partial_{r}\left[\frac{1}{2\left(br\right)^{d}}\left(\vphantom{\frac{\sigma_{\alpha\beta}\sigma^{\alpha\beta}}{d-1}}r^{2}\left(1-\left(br\right)^{d}\right)\right.\right.\nonumber \\
 &  & \left.\left.-\frac{1}{2}\omega_{\alpha\beta}\omega^{\alpha\beta}-\left(br\right)^{2}K_{2}\left(br\right)\frac{\sigma_{\alpha\beta}\sigma^{\alpha\beta}}{d-1}\right)\vphantom{\frac{1}{2\left(br\right)^{d}}}\right]\nonumber \\
 &  & -\frac{1}{2r}u^{\lambda}\partial_{\lambda}u_{\sigma}P_{\mu}^{\lambda}\partial_{\lambda}u^{\sigma}+br\partial_{r}F\left(br\right)u^{\lambda}\partial_{\lambda}u_{\sigma}\sigma_{\mu}^{\sigma}\nonumber \\
 &  & +\frac{1}{r}\frac{2}{d-2}P_{\mu}^{\rho}\left(\partial_{\lambda}-\left(d-3\right)A_{\lambda}\right)\omega_{\rho}^{\lambda}\nonumber \\
 &  & +\left(\frac{2L\left(br\right)}{r\left(br\right)^{d-2}}-\frac{1}{2}\partial_{r}\left(\frac{2L\left(br\right)}{\left(br\right)^{d-2}}\right)\right.\nonumber \\
 &  & \left.+\frac{1}{r}\frac{1}{d-2}\vphantom{\frac{2L\left(br\right)}{\left(br\right)^{d-2}}}\right)P_{\mu}^{\rho}\left(\partial_{\lambda}-\left(d-1\right)A_{\lambda}\right)\sigma_{\rho}^{\lambda}\nonumber \\
 &  & +\frac{1}{2r}u^{\lambda}\partial_{\lambda}u^{\sigma}\partial_{\sigma}u_{\mu}\nonumber \\
 & = & -\Gamma_{r\mu}^{r\left(0+1+2\right)}\end{eqnarray}
}{\large \par}

{\large \begin{equation}
\ell^{\mu}\Gamma_{r\mu}^{\sigma\left(0+1+2\right)}\ell_{\sigma}=0\end{equation}
}{\large \par}

\chapter{{\LARGE The Constarint Equations to The Third Order}}

{\large We will present here a different method than what we presented
in chapter 6 to find the constraint equations to the third order of
the event and apparent horizons. }{\large \par}

{\large In order to do so, we look at equation (\ref{eq:Claim}) and
use (\ref{eq:getting rid of the metric}) to get the following form
of the constraint equations:}{\large \par}

{\large \begin{equation}
R_{ab}m^{a}e_{\mu}^{b}=0\end{equation}
}{\large \par}

{\large In order to check that we get the same constraint equations
from the two hyper-surfaces, we calculate the difference between the
constraint equations evaluated on the event horizon and the constraint
equations evaluated on the apparent horizon, to the third order in
derivative expansion, i.e.:}{\large \par}

{\large \begin{eqnarray}
EH-AH & \equiv & \left.R_{ab}m^{a}e_{\mu}^{b}\right|_{r=r_{EH}}-\left.R_{ab}m^{a}e_{\mu}^{b}\right|_{r=r_{AH}}\label{eq:Difference between EH and AH}\end{eqnarray}
}{\large \par}

{\large The quantities will be calculated from the full second order
bulk metric (\ref{eq:general metric}). Bearing in mind that the difference
in $m^{a}$ only starts from the second order, and that the difference
in $R_{ab}$ and $e_{\mu}^{a}$ only starts from the third order,
we can write (\ref{eq:Difference between EH and AH}) in the following
manner: \begin{eqnarray}
EH-AH & = & \left(\left.R_{a\mu}^{\left(3\right)}\right|_{r=r_{EH}}-\left.R_{a\mu}^{\left(3\right)}\right|_{r=r_{AH}}\right)m^{a\left(0\right)}\nonumber \\
 &  & +\left(\left.m^{a\left(2\right)}\right|_{r=r_{EH}}-\left.m^{a\left(2\right)}\right|_{r=r_{AH}}\right)R_{a\mu}^{\left(1\right)}+\nonumber \\
 &  & +\left(\left.e_{\mu}^{b\left(3\right)}\right|_{r=r_{EH}}-\left.e_{\mu}^{b\left(3\right)}\right|_{r=r_{AH}}\right)R_{ab}^{\left(0\right)}m^{a\left(0\right)}\nonumber \\
 &  & +\left(\left.m^{a\left(3\right)}\right|_{r=r_{EH}}-\left.m^{a\left(3\right)}\right|_{r=r_{AH}}\right)R_{a\mu}^{\left(0\right)}+\nonumber \\
 &  & +\left(\left.m^{a\left(2\right)}\right|_{r=r_{EH}}-\left.m^{a\left(2\right)}\right|_{r=r_{AH}}\right)R_{ab}^{\left(0\right)}e_{\mu}^{b\left(1\right)}+\nonumber \\
 &  & +\left(\left.R_{ab}^{\left(2\right)}\right|_{r=r_{EH}}-\left.R_{ab}^{\left(2\right)}\right|_{r=r_{AH}}\right)m^{a\left(0\right)}e_{\mu}^{b\left(1\right)}\nonumber \\
 &  & +\left(\left.R_{a\mu}^{\left(2\right)}\right|_{r=r_{EH}}-\left.R_{a\mu}^{\left(2\right)}\right|_{r=r_{AH}}\right)m^{a\left(1\right)}\label{eq:Difference between EH and AH detailed}\end{eqnarray}
where the superscript refers to the order that we are evaluating the
specific quantities, i.e., if it is written $^{\left(1\right)}$ only
the first derivative terms should be taken (no zero terms). The quantities
in (\ref{eq:Difference between EH and AH detailed}) that are outside
the brackets should be evaluated at the location $r=b^{-1},$ for
instance:}{\large \par}

{\large \begin{equation}
\left.m^{r\left(0\right)}\right|_{r=b^{-1}}=\left.g^{rr\left(0\right)}\right|_{r=b^{-1}}=\left.r^{2}f\left(br\right)\right|_{r=b^{-1}}=0\end{equation}
}{\large \par}

{\large Here we write some of the quantities that will help us to
evaluate (\ref{eq:Difference between EH and AH detailed}),}{\large \par}

{\large \begin{eqnarray}
 &  & R_{rr}^{\left(0\right)}=0,\nonumber \\
 &  & m^{r\left(0\right)}=0,\quad m^{\nu\left(1\right)}=0,\nonumber \\
 &  & \left.m^{\mu\left(2\right)}\right|_{r=r_{EH}}=\left.m^{\mu\left(2\right)}\right|_{r=r_{AH}},\label{eq:Identities for the difference calculation}\\
 &  & R_{r\nu}^{\left(0\right)}=du_{\nu},\quad R_{r\nu}^{\left(1\right)}=R_{r\nu}^{\left(2\right)}=R_{r\nu}^{\left(3\right)}=0,\nonumber \\
 &  & e_{\mu EH/AH}^{a\left(1\right)}=e_{\mu EH/AH}^{r\left(1\right)}=\partial_{\mu}r_{EH/AH}^{\left(0\right)}=\partial_{\mu}b^{-1},\quad e_{\mu}^{a\left(3\right)}=e_{\mu}^{r\left(3\right)}=\partial_{\mu}r^{\left(2\right)}\nonumber \end{eqnarray}
}{\large \par}

{\large We know that only the third order terms that will be different
are the one that are composed of the second order horizon location
$r^{\left(2\right)}$, which is different in both hyper-surfaces.
For instance, the only part that will contribute in the third order
Ricci tensor, is the Ricci tensor calculated to the first order in
derivatives evaluated on one of the hyper-surfaces and from this we
will take the third order terms that come from $r_{EH/AH}^{\left(2\right)}$.
We denote $\triangleq$ to represent the only terms that will not
cancel out in the difference $\left.m^{a\left(3\right)}\right|_{r=r_{EH}}-\left.m^{a\left(3\right)}\right|_{r=r_{AH}}$,
for example we will write,}{\large \par}

{\large \begin{eqnarray}
\left.m^{\mu\left(3\right)}\right|_{r=r_{EH/AH}} & = & g^{\mu r\left(3\right)}-g^{\mu\nu\left(0\right)}\partial_{\nu}\left(r_{EH/AH}^{\left(2\right)}\left(x\right)\right)-g^{\mu\nu\left(2\right)}\partial_{\nu}\left(r^{\left(0\right)}\left(x\right)\right)\nonumber \\
 & \triangleq & -\frac{1}{r}u^{\lambda}\partial_{\lambda}u^{\mu}-\frac{1}{r^{2}}P^{\mu\nu}\partial_{\mu}r_{EH/AH}^{\left(2\right)}-\frac{1}{r^{2}}P^{\mu\nu}\partial_{\nu}b^{-1}\nonumber \\
 & = & -\frac{1}{r}u^{\lambda}\partial_{\lambda}u^{\mu}+\frac{1}{r^{2}}b^{-1}u^{\lambda}\partial_{\lambda}u^{\mu}-b^{2}P^{\mu\nu}\partial_{\mu}r_{EH/AH}^{\left(2\right)}\nonumber \\
 & \triangleq & -b^{2}r_{EH/AH}^{\left(2\right)}u^{\lambda}\partial_{\lambda}u^{\mu}-b^{2}P^{\mu\nu}\partial_{\mu}r_{EH/AH}^{\left(2\right)}\end{eqnarray}
}{\large \par}

{\large \begin{eqnarray}
\left.m^{r\left(3\right)}\right|_{r=r_{EH/AH}} & = & g^{rr\left(3\right)}-g^{r\nu\left(0\right)}\partial_{\nu}\left(r_{EH/AH}^{\left(2\right)}\left(x\right)\right)-g^{r\nu\left(2\right)}\partial_{\nu}\left(r^{\left(0\right)}\left(x\right)\right)\nonumber \\
 & \triangleq & -2r\frac{\partial_{\lambda}u^{\lambda}}{d-1}-u^{\nu}\partial_{\nu}r_{EH/AH}^{\left(2\right)}\nonumber \\
 & \triangleq & -2r_{EH/AH}^{\left(2\right)}\frac{\partial_{\lambda}u^{\lambda}}{d-1}-u^{\nu}\partial_{\nu}r_{EH/AH}^{\left(2\right)}\end{eqnarray}
where $r_{EH/AH}$ is the event or apparent horizon full location
up to the second order, and $r^{\left(0\right)}=b^{-1}$.}{\large \par}

{\large Now we are ready to evaluate each term in (\ref{eq:Difference between EH and AH detailed}):}{\large \par}

{\large \begin{eqnarray*}
 &  & \left(\left.R_{a\mu}^{\left(2\right)}\right|_{r=r_{EH}}-\left.R_{a\mu}^{\left(2\right)}\right|_{r=r_{AH}}\right)m^{a\left(1\right)}=\\
 &  & =\left(\left.R_{r\mu}^{\left(2\right)}\right|_{r=r_{EH}}-\left.R_{r\mu}^{\left(2\right)}\right|_{r=r_{AH}}\right)m^{r\left(1\right)}+\left(\left.R_{\nu\mu}^{\left(2\right)}\right|_{r=r_{EH}}-\left.R_{\nu\mu}^{\left(2\right)}\right|_{r=r_{AH}}\right)m^{\nu\left(1\right)}\\
 &  & =0\end{eqnarray*}
\begin{eqnarray*}
 &  & \left(\left.R_{ab}^{\left(2\right)}\right|_{r=r_{EH}}-\left.R_{ab}^{\left(2\right)}\right|_{r=r_{AH}}\right)m^{a\left(0\right)}e_{\mu}^{b\left(1\right)}=\\
 &  & =\left(\left.R_{rr}^{\left(2\right)}\right|_{r=r_{EH}}-\left.R_{rr}^{\left(2\right)}\right|_{r=r_{AH}}\right)m^{r\left(0\right)}e_{\mu}^{r\left(1\right)}+\left(\left.R_{\mu r}^{\left(2\right)}\right|_{r=r_{EH}}-\left.R_{\nu r}^{\left(2\right)}\right|_{r=r_{AH}}\right)m^{\nu\left(0\right)}e_{\mu}^{r\left(1\right)}\\
 &  & =0\end{eqnarray*}
}{\large \par}

{\large \begin{eqnarray*}
 &  & \left(\left.m^{a\left(2\right)}\right|_{r=r_{EH}}-\left.m^{a\left(2\right)}\right|_{r=r_{AH}}\right)R_{ab}^{\left(0\right)}e_{\mu}^{b\left(1\right)}=\\
 &  & =\left(\left.m^{r\left(2\right)}\right|_{r=r_{EH}}-\left.m^{r\left(2\right)}\right|_{r=r_{AH}}\right)R_{rr}^{\left(0\right)}e_{\mu}^{r\left(1\right)}+\left(\left.m^{\nu\left(2\right)}\right|_{r=r_{EH}}-\left.m^{\nu\left(2\right)}\right|_{r=r_{AH}}\right)R_{\nu r}^{\left(0\right)}e_{\mu}^{r\left(1\right)}\\
 &  & =0\end{eqnarray*}
}{\large \par}

{\large \begin{eqnarray*}
 &  & \left(\left.m^{a\left(3\right)}\right|_{r=r_{EH}}-\left.m^{a\left(3\right)}\right|_{r=r_{AH}}\right)R_{a\mu}^{\left(0\right)}=\\
 &  & =\left(\left.m^{r\left(3\right)}\right|_{r=r_{EH}}-\left.m^{r\left(3\right)}\right|_{r=r_{AH}}\right)R_{r\mu}^{\left(0\right)}+\left(\left.m^{\nu\left(3\right)}\right|_{r=r_{EH}}-\left.m^{\nu\left(3\right)}\right|_{r=r_{AH}}\right)R_{\nu\mu}^{\left(0\right)}\\
 &  & =-d\left(\left.r^{\left(2\right)}\right|_{r=r_{EH}}-\left.r^{\left(2\right)}\right|_{r=r_{AH}}\right)2\frac{\partial_{\lambda}u^{\lambda}}{d-1}u_{\mu}\\
 &  & +db^{-2}P_{\mu\nu}\left(\left.r^{\left(2\right)}\right|_{r=r_{EH}}-\left.r^{\left(2\right)}\right|_{r=r_{AH}}\right)b^{2}u^{\lambda}\partial_{\lambda}u^{\nu}\\
 &  & -du^{\nu}u_{\mu}\left(\left.\partial_{\nu}\left(r^{\left(2\right)}\left(x\right)\right)\right|_{r=r_{EH}}-\left.\partial_{\nu}\left(r^{\left(2\right)}\left(x\right)\right)\right|_{r=r_{AH}}\right)\\
 &  & +dP_{\mu}^{\sigma}\left(\left.\partial_{\sigma}\left(r^{\left(2\right)}\left(x\right)\right)\right|_{r=r_{EH}}-\left.\partial_{\sigma}\left(r^{\left(2\right)}\left(x\right)\right)\right|_{r=r_{AH}}\right)\\
 &  & =-d\left(\left.r^{\left(2\right)}\right|_{r=r_{EH}}-\left.r^{\left(2\right)}\right|_{r=r_{AH}}\right)\left(2\frac{\partial_{\lambda}u^{\lambda}}{d-1}u_{\mu}-u^{\lambda}\partial_{\lambda}u_{\nu}\right)\\
 &  & +d\left(\left.\partial_{\mu}\left(r^{\left(2\right)}\left(x\right)\right)\right|_{r=r_{EH}}-\left.\partial_{\mu}\left(r^{\left(2\right)}\left(x\right)\right)\right|_{r=r_{AH}}\right)\end{eqnarray*}
}{\large \par}

{\large \begin{eqnarray*}
 &  & \left(\left.e_{\mu}^{b\left(3\right)}\right|_{r=r_{EH}}-\left.e_{\mu}^{b\left(3\right)}\right|_{r=r_{AH}}\right)R_{ab}^{\left(0\right)}m^{a\left(0\right)}=\\
 &  & =\left(\left.e_{\mu}^{r\left(3\right)}\right|_{r=r_{EH}}-\left.e_{\mu}^{r\left(3\right)}\right|_{r=r_{AH}}\right)R_{rr}^{\left(0\right)}m^{r\left(0\right)}+\left(\left.e_{\mu}^{r\left(3\right)}\right|_{r=r_{EH}}-\left.e_{\mu}^{r\left(3\right)}\right|_{r=r_{AH}}\right)R_{r\nu}^{\left(0\right)}m^{\nu\left(0\right)}\\
 &  & =-d\left(\left.\partial_{\mu}\left(r^{\left(2\right)}\left(x\right)\right)\right|_{r=r_{EH}}-\left.\partial_{\mu}\left(r^{\left(2\right)}\left(x\right)\right)\right|_{r=r_{AH}}\right)\end{eqnarray*}
\begin{eqnarray*}
 &  & \left(\left.m^{a\left(2\right)}\right|_{r=r_{EH}}-\left.m^{a\left(2\right)}\right|_{r=r_{AH}}\right)R_{a\mu}^{\left(1\right)}=\\
 &  & =\left(\left.m^{\nu\left(2\right)}\right|_{r=r_{EH}}-\left.m^{\nu\left(2\right)}\right|_{r=r_{AH}}\right)R_{\nu\mu}^{\left(1\right)}+\left(\left.m^{r\left(2\right)}\right|_{r=r_{EH}}-\left.m^{r\left(2\right)}\right|_{r=r_{AH}}\right)R_{r\mu}^{\left(1\right)}=0\end{eqnarray*}
}{\large \par}

{\large \begin{eqnarray*}
 &  & \left(\left.R_{a\mu}^{\left(3\right)}\right|_{r=r_{EH}}-\left.R_{a\mu}^{\left(3\right)}\right|_{r=r_{AH}}\right)m^{a\left(0\right)}=\\
 &  & =\left(\left.R_{r\mu}^{\left(3\right)}\right|_{r=r_{EH}}-\left.R_{r\mu}^{\left(3\right)}\right|_{r=r_{AH}}\right)m^{r\left(0\right)}+\left(\left.R_{\nu\mu}^{\left(3\right)}\right|_{r=r_{EH}}-\left.R_{\nu\mu}^{\left(3\right)}\right|_{r=r_{AH}}\right)m^{\nu\left(0\right)}\\
 &  & =\left(\left.R_{\nu\mu}^{\left(3\right)}\right|_{r=r_{EH}}-\left.R_{\nu\mu}^{\left(3\right)}\right|_{r=r_{AH}}\right)u^{\nu}\\
 &  & =d\left(\left.r^{\left(2\right)}\right|_{r=r_{EH}}-\left.r^{\left(2\right)}\right|_{r=r_{AH}}\right)\left(2\frac{\partial_{\lambda}u^{\lambda}}{d-1}u_{\mu}-u^{\lambda}\partial_{\lambda}u_{\nu}\right)\end{eqnarray*}
}{\large \par}

{\large We seek the third order constraint equations. This is the
reason that in our calculation we used the previous constraint equations
(\ref{eq:constraint equation 1st order 1}), (\ref{eq:constraint equations 1st order 2})
that are of the second order. Assembling all of the above terms and
inserting them into (\ref{eq:Difference between EH and AH detailed}),
reveals that our statement is correct, for we found:}{\large \par}

{\large \begin{eqnarray}
EH-AH & \equiv & \left.R_{ab}m^{a}e_{\mu}^{b}\right|_{r=r_{EH}}-\left.R_{ab}m^{a}e_{\mu}^{b}\right|_{r=r_{AH}}=0\label{eq:Difference between EH and AH=00003D0}\end{eqnarray}
}{\large \par}

\end{document}